\documentclass[sigconf,10pt]{acmart}

\usepackage{amsmath,amssymb}
\usepackage{amsfonts}
\usepackage{amsthm}
\usepackage{mathtools}
\usepackage{graphicx}
\usepackage{xcolor}
\usepackage{balance}  % for  \balance command ON LAST PAGE  (only there!)
\usepackage{pbox}
\usepackage{color,soul}
\usepackage{colortbl}
\usepackage{tabularx}
\usepackage{subcaption}
\usepackage{caption}
\usepackage{arydshln}
\usepackage[plain]{algorithm}
\usepackage[noend]{algpseudocode}
\usepackage{lipsum}
\usepackage{multirow}

\DeclareTextSymbolDefault{\textsterling}{T1}

\copyrightyear{}
\acmYear{}
\acmDOI{}

\newcommand{\system}{Daisy~}
\newcommand{\systemNoindent}{Daisy}

\newtheorem{example}{Example}%[section]
\newtheorem{definition}{Definition}%[section]
\newtheorem{lemma}{Lemma}%[section]

\definecolor{OliveGreen}{HTML}{556B2F} 

\begin{document}	
	
	\fancyhead{} 	
\title{Cleaning Denial Constraint Violations through Relaxation}

\author{Stella Giannakopoulou}
\affiliation{%
	\institution{EPFL}
}
\email{stella.giannakopoulou@epfl.ch}

\author{Manos Karpathiotakis}
\authornote{Work done while the author was at EPFL.}
\affiliation{%
	\institution{Facebook}
}
	\email{manos@fb.com}
\author{Anastasia Ailamaki}
\affiliation{%
	\institution{EPFL}
}
\email{anastasia.ailamaki@epfl.ch}

	\renewcommand{\shortauthors}{Stella Giannakopoulou, Manos Karpathiotakis, Anastasia Ailamaki}
	
	\begin{abstract}
	\vspace{-0.3em}
	
Data cleaning is a time-consuming process that depends on the data analysis that users perform. 
Existing solutions treat data cleaning as a separate offline process that takes place before analysis begins.
Applying data cleaning before analysis assumes a priori knowledge of the
inconsistencies and the query workload, thereby requiring effort on understanding and cleaning the data that is unnecessary for the analysis.
%Thus, users need to invest most of their time to curate dirty data
%before executing analysis tasks.
 
We propose an approach that performs probabilistic repair of denial constraint violations
on-demand, driven by the exploratory analysis that users perform.
We introduce \systemNoindent, a system that
seamlessly integrates data cleaning into the analysis by relaxing query results.
\system executes analytical query-workloads over dirty data by weaving
cleaning operators into the query plan.
Our evaluation shows that \system adapts to the workload and outperforms traditional offline cleaning on both synthetic and real-world
workloads.
%iii) guarantees correctness for functional dependencies and provides accuracy guarantees
%for general denial constraints.

%This paper introduces \systemNoindent, a system that
%integrates data cleaning tasks seamlessly into the data analysis process.
%\system executes analytical query workloads over dirty data by weaving
%cleaning operators into the query plan.
%We implement \system over Spark and show that it scales better than existing
%offline data cleaning approaches.
\end{abstract}

	%\begin{CCSXML}
	%<ccs2012>
	%  <concept_id>10010520.10010553.10010562</concept_id>
	%  <concept_desc>Computer systems organization~Embedded systems</concept_desc>
	%  <concept_significance>500</concept_significance>
	%  <concept_id>10010520.10010575.10010755</concept_id>
	%  <concept_desc>Computer systems organization~Redundancy</concept_desc>
	%  <concept_significance>300</concept_significance>
	%  <concept_id>10010520.10010553.10010554</concept_id>
	%  <concept_desc>Computer systems organization~Robotics</concept_desc>
	%  <concept_significance>100</concept_significance>
	%  <concept_id>10003033.10003083.10003095</concept_id>
	%  <concept_desc>Networks~Network reliability</concept_desc>
	%  <concept_significance>100</concept_significance>
	%</ccs2012>
	%\end{CCSXML}
	%
	%\ccsdesc[500]{Computer systems organization~Embedded systems}
	
	
	
	\maketitle

\section{Introduction}
\label{sec:intro}

% General intro -- needs to address the exploratory analysis
%Data scientists collect and analyze data in order to extract valuable information out of it. 
Real-life data contain erroneous information, which leads to inaccurate data analysis~\cite{incomplete_dbs,activeclean}.
Data scientists spend most of their time cleaning their data~\cite{buckley_new_2012}, until they are able to extract insights.
Depending on the accuracy requirements of the workload and the data they need to access, users iteratively apply cleaning tasks
until they are satisfied with the resulting quality.
%spend most of their time on cleaning and organizing their data~\cite{buckley_new_2012} . 
Thus, data cleaning is a subjective and time-consuming process.

Data cleaning is an interactive and exploratory process that involves expensive operations. 
Error detection requires multiple pairwise comparisons to check the satisfiability of the rules \cite{cleanm}.
Data repairing adds an extra overhead as it requires many iterations of assigning candidate values to
dirty cells, until all rules are satisfied \cite{holistic_dc, nadeef}. 
%At the same time, data cleaning depends on the analysis that users perform;
Data scientists also detect inconsistencies and constraints 
at data exploration time~\cite{quERy}. 
Hence, traversing the whole dataset multiple times to repair each discovered discrepancy is cost-prohibitive.

State-of-the-art approaches can be divided into offline, and online analysis-aware approaches. 
Offline tools \cite{nadeef,bigdansing, holoclean} treat data cleaning as a separate process, decoupled from analysis. Applying data cleaning before analysis begins
requires prior knowledge of the errors that exist. Offline cleaning is also cost-prohibitive, as it
operates over the whole dataset \cite{data_quality}. Analysis-aware tools \cite{quERy,sampleclean,activeclean,qo_dynamic_imputation} focus on entity resolution or deduplication, or they limit themselves to cell-level errors.
But entity resolution tools either require expensive preprocessing \cite{quERy} or support only aggregate queries~\cite{sampleclean}.

%limit themselves to entity resolution~\cite{entity_resolution},
%while still requiring an expensive preprocessing blocking phase.
%disregarding the existence of data cleaning issues when executing ad-hoc analysis tasks, brings inaccuracies to the information that users extract \cite{activeclean}. 

% Say what is needed ?
%
There is a need for an efficient cleaning approach that is weaved into the exploratory analysis and that cleans data on-demand.
On-the-fly cleaning repairs only necessary data, thus if only a subset is analyzed, the wasted-effort is minimized.
Online cleaning also benefits offline cleaning by enhancing the predictability on the required cleaning tasks.
Thus, integrating cleaning with analysis efficiently supports exploratory applications \cite{exploratory_analysis} by
reducing data-to-insight time.

\begin{table}[t]
	\footnotesize
	\begin{center}
	\begin{tabular}{|c|c|c|}
		\hline
		\textbf{Name} & \textbf{Zip} & \textbf{City}\\
		\hline
		Jon & 9001 & Los Angeles\\
		\hline
		Jim & 9001  & San Francisco\\
		\hline
		Mary & 10001  & New York\\
		\hline
		Jane & 10002 & New York\\
		\hline   
	\end{tabular}	
%	\vspace{-1.em}
	\caption{Employees dataset.}
	\label{tab:employees}
	\vspace{-3em}
	\end{center}
\end{table}

\vspace{-0.7em}
{\setlength{\parindent}{0em}
\begin{example}
\label{ex:intro}
	Consider the dataset of Table \ref{tab:employees} that comprises employees information. Assume that a user is interested in 
	analyzing all employees in Los Angeles. The insights that the user extracts might be incorrect
	due to the conflict among the first two tuples that have the same zip code and different city name; they violate the functional dependency \textit{zip$\rightarrow$city} stating that the zip code defines the city.
	Hence, the analysis ignores the second tuple whose city is San Francisco, but after cleaning it, it might obtain the value Los Angeles.
%	
%	by extracting the Los Angeles subset, the analysis
%	ignores the second tuple whose city might be Los Angeles after cleaning it. 
	Having to clean the whole dataset
	is unnecessary as (i) the user is interested only in a subset of the data, and (ii) the query result can be cleaned by checking only the relevant data subset.
\end{example}}
\vspace{-0.5em}

We present the first approach that intermingles cleaning denial constraint violations
with exploratory SPJ (Select-Project-Join) and aggregate queries, and that gradually cleans the data.
Denial constraints (DCs) comprise a family of rules that have been widely used to capture real-life data inconsistencies \cite{data_quality_srivastava,data_quality}.
%By coupling data cleaning with data analysis, we reduce the data cleaning cost by focusing only on 
%the interesting part of the dataset, that is, the part that the user accesses through the exploratory analysis process.
To provide correct results over dirty data, we introduce cleaning operators in the query plan
and employ a cost model to optimally place them.
To enable cleaning operators to detect errors, we define at the execution level a novel query-result relaxation mechanism in the context of 
DCs.
Query-result relaxation enhances the query result with correlated data from the dataset to allow
error detection.
Then, given the detected errors, we propose candidate fixes by providing probabilistic results \cite{probabilistic_dbs}.
%After query execution, we isolate the changes and apply the delta to the original dataset.
%Thus, we incrementally clean the data by adding an overhead to each query. 
%In addition, we build a
%cost model which decides between incremental cleaning and full cleaning by taking into consideration the type of
%queries, the number of violations of the dataset, and the accuracy of the resulting answer. 
We validate our approach by building
\systemNoindent, a distributed incremental cleaning framework over Spark \cite{spark}.
%The long version of the paper can be found in \cite{incremental_cleaning}.

\noindent\textbf{Contributions:} Our contributions are as follows:
\vspace{-0.6em}
\begin{itemize}%[leftmargin=*]	
	\item We present a query result relaxation mechanism that enables	
	interleaving SPJ and aggregate queries with cleaning DC violations.
	Our approach guarantees correctness, compared to the offline approach in the case of functional dependencies,
	and provides accuracy estimates in the presence of general DCs.
%	Query answer relaxation enhances query answers with correlated entities from the dataset 
%	which allow fixing the errors, thereby including all candidate tuples
%	in the answer.
	
	\item We introduce cleaning operators inside the query plan by using a cost model that
	determines the execution order of cleaning and query operators at query time.
%	To decide on the
%	order, the cost model takes into consideration information such as the type of rules, the type of query, and the number of violations.
%	
	\item We implement \systemNoindent, the first system that enables exploratory data-analysis queries over data with DC violations.
	We execute \system over Spark, and experimentally show
	that it is faster than offline cleaning solutions on synthetic data and  
	supports real-world workloads that offline cleaning fails to address. 
	\vspace{-0.5em}
\end{itemize}

%\noindent The structure of the paper is organized as follows. 
%Section~\ref{sec:related} presents related work.
%Section~\ref{sec:preliminaries} presents the background on offline cleaning.
%Section~\ref{sec:query_relax} presents query execution over dirty data. 
%Section~\ref{sec:loglevel_optimizations} presents the cleaning-aware logical plan.
%%Section~\ref{sec:accuracy} shows the accuracy of incremental cleaning.
%Section~\ref{sec:system} presents the architecture of \systemNoindent.
%Section~\ref{sec:eval} evaluates \systemNoindent.
%Finally, Section~\ref{sec:conclusion} concludes the paper.
\section{Related Work}
\label{sec:related}

%Data cleaning is still a data management challenge~\cite{error_detection,cleaning_challenges,datacleaning_survey}.
In this section we survey related work and highlight how our work advances the state of the art.

\noindent\textbf{Offline Integrity Constraint Tools}: 
To repair denial constraint errors, offline systems follow a centralized or a distributed approach.
%NADEEF and LLUNATIC~\cite{nadeef,llunatic} tackle the problem of cleaning denial constraint violations. %, and tries to update erroneous values in a way that all the rules are satisfied \cite{holistic_dc}.
%
%BigDansing~\cite{bigdansing} ports the insights of NADEEF 
%in a distributed setting by extending MapReduce-like 
%frameworks with support for duplicate elimination and denial constraints.
NADEEF~\cite{nadeef} assumes known candidate fixes. BigDansing \cite{bigdansing} 
introduces and optimizes logical cleaning operators.
For repairing, it provides suggestions comprising the condition between the dirty cells or 
blindly assigns values until all rules are satisfied \cite{holistic_dc}. 
LLUNATIC \cite{llunatic} repairs data by using the principle of minimality.
\system differs as it
provides probabilistic candidate fixes for each dirty cell.

Holoclean \cite{holoclean} repairs data by combining integrity constraints, master data, and quantitative statistics.
%To combine different signals, Holoclean uses an extension of Datalog which captures probabilistic distributions. 
%Then for the inference, it starts grounding the rules based on the domain of each erroneous cell. 
%For scalability purposes, Holoclean limits the domain of each cell by taking into consideration the co-occurrences of the erroneous cells with the values of other cells, and combines tuples that participate in
%detected violations.
Holoclean differs from \system in that it relies on master data and on training based on the clean part of the dataset. \system
uses the provided dependencies and computes probabilistic candidate fixes for the erroneous entities.

NADEEF, BigDansing and Holoclean differ from \system in that they are offline data cleaning systems that operate over the whole dataset, before data analysis begins.

%\begin{sloppypar}
\noindent\textbf{Online, Analysis-aware Cleaning}: \texttt{QuERy} \cite{quERy} intermingles deduplication with query processing.
\texttt{QuERy} uses blocking \cite{blocking} for preprocessing
%which groups the entities into clusters of similar entities 
and introduces operators in the query plan, which operate
over the blocks. \texttt{QuERy} also optimizes the plan that involves cleaning operators. %\texttt{QuERy} adds the cleaning operators in the query plan using two approaches: a) a lazy approach that defers cleaning until it is necessary, and b) an adaptive approach 
%i) a lazy approach in which it defers the cleaning by estimating the blocks which are candidates to satisfy the query, and ii) an adaptive approach, which decides on the optimal placement of the cleaning operators in the query tree.
%\end{sloppypar}
SampleClean~\cite{sampleclean} extracts a sample out of a dataset with duplicates and cell-level errors, asks users to clean it, and uses the sample to answer aggregate queries. SampleClean estimates the query result
given the cleaned data and corrects the error of the queries over the uncleaned data.
\texttt{QuERy} and SampleClean address entity resolution, duplicates, or cell-level errors, whereas \system focuses on integrity constraints.
Also, \texttt{QuERy} differs in that it requires preprocessing to apply the blocking.
Finally, SampleClean supports only aggregate queries.
ActiveClean~\cite{activeclean} incrementally updates a machine-learning model as the user cleans the data.
ActiveClean addresses cell-level corruption cases, excluding cases that involve
multiple records such as integrity constraints.
ImputeDB~\cite{qo_dynamic_imputation} considers query processing over data with missing values and decides whether to drop tuples by choosing the 
optimal solution in the efficiency/quality trade-off. 
ImputeDB also limits itself to cell-level errors.

\noindent\textbf{Consistent Query Answering}: The area of consistent query answering \cite{arenas_cqa,cavsat,cqa_ib,cqa_fo_rewrite} focuses on computing the query answers that are consistent with all possible repairs without modifying the data.
%given a set of integrity constraints, 
%and on providing query answers, by using the tuples that belong to every repair.
%These repairs constitute datasets which are minimally different from the erroneous dataset by adding or deleting entities. 
\system differs in that it computes all candidate qualifying tuples, and it applies repairing incrementally,
driven by the queries and the denial constraints.

%\textbf{Query Relaxation}:

\section{From Offline to \\ Online Data cleaning}
\label{sec:preliminaries}

%In this section we present the problem and the traditional offline approach for data cleaning.
%, and the motivation for integrating cleaning tasks with query processing.

%\noindent \textbf{Problem Statement}.
%\noindent\textbf{Cleaning violations of denial constraints (DCs)}.
%\system addresses the efficiency problem of detecting and repairing violations of DCs in exploratory analysis scenarios.
\noindent
	\textbf{Problem Statement:} 
	%Traditional offline data cleaning is time consuming and depends on the analysis that users perform. 
	We need to efficiently clean in real-time exploratory query results in the presence of dirty data. 
	We clean
	denial constraint (DC) violations~\cite{data_quality} as they involve
	a wide range of rules that detect semantic inconsistencies in the data.
	DCs are universally quantified first-order logic sentences 
	that represent data dependencies, including functional dependencies (FDs). 
	DCs are defined as:
	%\vspace{-0.5em}
	{
		%\[\forall t_1, ...,t_k \neg (p(x_1) \land p(x_2)  \land ... p(x_n))\] 
		$\forall$$t_1$, ...,$t_k$$\neg$($p_{1}$$\land p_2...  \land p_m$),
	}
	where each $t_i$ is a tuple, each $p_i$ is a predicate involving conditions between the attributes of one or more tuples, $k$ is the number of involved tuples, and $m$ is the number of predicates.

\noindent\textbf{Challenges: } Interleaving cleaning with querying must provide accurate results,
	without cleaning the whole dataset. 
	Moreover, as cleaning is costly compared to query processing,
	adding the cleaning overhead over each query might result in an overall cost higher than offline cleaning.
	Also, during data exploration, users have partial knowledge on the rules that hold; cleaning a value, given
	partial knowledge affects the resulting data quality \cite{glitches}. 
	Even when the rules are known, automatically fixing an error might result in inaccuracies \cite{holoclean}; human effort or master data are required.

%Solution

\noindent\textbf{Solution: } 
	To efficiently and accurately provide correct query results in the presence of DCs, we weave cleaning operators into the query plan.
	We optimize the overall execution by detecting the relevant 
	data subset that affects the cleanliness of the result,
	and we introduce a cost model to optimally place 
	the cleaning operators, based on the overhead they add on each query.	
%	We show that our approach 
%	guarantees correctness, compared to offline cleaning for FDs, and we provide accuracy estimates for DCs.
	To capture partial knowledge of the rules and the data, we clean data by providing probabilistic fixes.	
	Then, using our solution once all rules are known and given the probabilistic suggestions, we can either use inference~\cite{scare,holoclean,optimal_repairs} when master data exist, or have
	humans fix the errors in the query results.
	Inference approaches over the probabilistic data are complementary and out of the scope of this work. 
%	In Section \ref{subsec:realworld} we show how existing inference approaches can be used over the probabilistic fixes using Holoclean \cite{holoclean}.
	Future work includes examining the human cost of 
	cleaning the flagged dirty values of the query results.

\vspace{-0.5em}
\section{Query Execution \\ over Dirty Data}
\label{sec:query_relax}
\vspace{-0.2em}

%%%%%%%%%%%%%%%%%%%%%%ADD%%%%%%%%%%%%%%%%%
Executing queries over dirty data induce wrong query results \cite{activeclean}. A tuple might erroneously appear or be missing
from a query result due to a dirty value. 
We describe how we fix wrong query-results, by detecting and cleaning potentially qualifying tuples. 
To detect qualifying tuples, we introduce the query-result relaxation mechanism that relaxes results, based on the dependencies
defined by the rules.  
%%%%%%%%%%%%%%%%%%%%%%ADD%%%%%%%%%%%%%%%%%
Query result relaxation differs from query relaxation \cite{queryrelax_incomplete} in that instead of relaxing the conditions of the query,
it relaxes the result. Still, we relax the result to compute conflicting tuples based on the input DCs, whereas query relaxation
is used to deal with failing queries and incomplete databases.
To clean the relaxed result, we provide probabilistic fixes
based on the frequency of each candidate value in a dirty cell.
Then, we update the dataset in-place with the probabilities. We also maintain provenance
to the original values in case new rules appear.
Thus, we gradually transform the dataset into a probabilistic dataset. 
%We maintain probabilistic values since in real-world exploratory scenarios
%users identify rules while exploring through the data; updating a value given partial
%knowledge of the rules affects the resulting data quality~\cite{glitches}.
%In the following, we present how we apply cleaning at query time by introducing cleaning operators. 

Our probabilistic representation uses attribute-level uncertainty~\cite{probabilistic_dbs}; attributes take
multiple candidate values. To represent candidate tuples (i.e., possible worlds) by using attribute-level representation, we store
in each candidate value an identifier of the possible world it belongs to. 
Then, query operators output a tuple iff at least one candidate value qualifies. 
%Thus, a value satisfies a condition if at least one candidate value qualifies.
Thus, (self-)joins on probabilistic join-keys output a pair iff
the candidate values of the join-keys overlap.
To enable reasoning over the data, each tuple of the result contains
all candidate values.
For (self-)joins, we also employ a similar approach to the lineage used in probabilistic data
\cite{probabilistic_dbs}; we store in the result the originating tuple IDs because if a potential inference
updates a join key value, a pair might no longer satisfy the join.
In the following, we introduce cleaning operators that enable cleaning at query time.

\vspace{-0.5em}
{\setlength{\parindent}{0em}
	\begin{definition}
		\label{def:cleaningop}
		A cleaning operator is an update operator that receives a query-result or a relation and outputs
		the clean result or relation. When the cleaning operator takes input from a query operator it
		(a) relaxes the result based on the dependencies of the input DCs,
		(b) detects and fixes errors, and (c) updates the data in-place with the clean values.
\end{definition}}
\vspace{-0.3em}
%The query answer relaxation mechanism first computes the dirty query answer, and then relaxes it by bringing extra candidate tuples 
%for satisfying the query. 
%The set of extra tuples that \system brings, consists of tuples which are similar to the ones belonging to the query answer;
%the similarity depends on the correlation that the tuples have with respect to the integrity constraints that hold in the dataset . 
%Cleaning operators provide correct answers by applying a query answer relaxation mechanism, which 
%enhances the dirty result with extra tuples from the dataset which allow the computation of the correct answer.
%After enhancing the query answer, the operator detects errors and computes the set of
%candidate values for each erroneous cell.
Cleaning operators differentiate between Select and Join operators.
For group-by queries, cleaning takes place before the aggregation; to avoid grouping recomputation, we push down cleaning either
over any underlying select or join condition, or over the input relation.
%After relaxing the answer, cleaning operators provide candidate probabilistic fixes
%based on the frequency that each
%candidate value appears in the erroneous cells.  
%Then, they update the dataset in-place with the probabilities, while also maintaining provenance
%information to the original values in case new rules appear.
%Query execution over probabilistic data follows the semantics of query execution over probabilistic databases \cite{probabilistic_survey, aggregate_queries, probabilistic_dbs},
%that is, a probabilistic value satisfies a condition if at least one candidate value qualifies.
%Each candidate value is also accompanied by the probability of being a fix for that
%erroneous cell. \system computes the candidate values by taking into consideration the correlations between the attributes given the constraints \cite{scare}.
%In the case of inequality predicates we exploit the partitioning of the cartesian product to restrict the number and the size
%of the partitions that need to be checked \TODO{recheck}.
Below, we present our probabilistic cleaning approach for SP and SPJ queries, given one or more DCs using relaxation.

%Apart from the type of query, the execution depends on the number of rules that hold.

%We introduce two cleaning operators depending on the underlying query operators in the logical plan:
%We use $clean_\sigma$ for the case of select queries, and
%$clean_{\Join}$ for join queries. The operator  $clean_\sigma$ gets
%input from a single relation, whereas  $clean_{\Join}$ gets input from two relations.
%Specifically, $clean_\sigma$ gets as input the output of a \textit{select} operator, it enriches
%the result using extra tuples from the dataset, then cleans the
%qualifying part, and updates the dataset with the modified tuples.
%Similarly, $clean_{\Join}$ gets the output of a join, and extracts
%the qualifying parts from both tables. Afterwards, $clean_{\Join}$
%cleans the corresponding erroneous tuples separately in each relation, updates the join result
%and re-checks for violations. 
%Thus, $clean_\Join$ might 
%changes are required in the join result.

%For example, assuming the rule
%$\forall t1, t2: \urcorner(t1.zip=t2.zip \wedge t1.City \neq t2.City)$ and the dataset shown in Table \ref{tab:dirty_cities}, then
%the city value of the first three tuples will be replaced by the probabilistic value $\{Los\ Angeles\, 67\%, San\ Francisco\, 33\%\}$.

\subsection{Cleaning SP query results given a FD}
\label{subsec:relax_fd}
\vspace{-0.2em}

%\vspace{-0.3em}
{\setlength{\parindent}{0em}
	\begin{definition}
		\label{def:sp}
		$clean_\sigma$ is a cleaning operator that relaxes and cleans the result of a \textit{select} operator.
\end{definition}}
\vspace{-0.3em}
%To clean the answer, $clean_{\sigma}$ applies a query answer relaxation mechanism 
%in order to identify the set of correlated tuples of the query answer. The correlated tuples 
%determine the candidate fixes of the erroneous values, since they contain values that
%appear together with the erroneous cells \cite{scare,holoclean}.
%The operator $clean_{\sigma}$ computes and cleans the relaxed result of the query.
%\subsubsection{SP Queries}
\noindent 
The first step of $clean_\sigma$ is to relax the result.
Consider a dataset with schema $S$, and a FD \textit{$\phi$: X$\rightarrow$Y}, where \textit{X$\subseteq$S}, \textit{Y$\subseteq$S}.
%denote a subset of the attributes of the dataset.
$X$ might contain multiple attributes, whereas \textit{Y} contains one attribute; if \textit{Y} contained
more attributes (e.g., $Y_1$,$Y_2$), then $\phi$ would be mapped to multiple FDs (e.g., \textit{$\phi_1$:X$\rightarrow$$Y_1$}, \textit{$\phi_2$:X$\rightarrow$$Y_2$}) \cite{datacleaning_survey}.
%In the following we describe the case of a single rule, and then we extend for multiple, potentially overlapping rules.
%\noindent\textbf{Single rule}: 
%\begin{definition}
Given a SP query with projection list \textit{P$\subseteq$S}, and where clause attributes W$\subseteq$S,  
$\phi$ affects query correctness iff (X$\cup$Y)$\cap$ (P$\cup$W) $\neq$$\emptyset$, i.e.,
%\end{definition}
iff the query accesses an attribute of $\phi$. 
If the query overlaps with $\phi$, $clean_\sigma$ augments the result with tuples
from the dataset that have the same value for $X$ and/or $Y$. We refer to the extra tuples as correlated tuples.
%The cleaning operator in the case of SP queries is defined as follows.
%Then, we trigger the violation detection over the augmented answer of the query.
%In the following, we distinguish between queries with filter conditions on the $lhs$ and the $rhs$ attributes respectively,
%since even though the $rhs$ values are determined by the value of the $lhs$, due to the dependency,
%the $lhs$ is non-deterministic. 

%\begin{minted}[escapeinside=||,fontsize=\footnotesize]{scala}
%def query_relax(query_answer, query q, constraints c)
%	lhs_set=
%\end{minted}

%\SetKwInOut{Parameter}{parameter}
%
%\begin{algorithm}
%	\small
%	\SetKwData{LHSSET}{lhs\_set}
%	\SetKwData{RHSSET}{rhs\_set}
%	\SetKwData{CORRELATED}{correlated\_tuples}
%	\SetKwData{TEMPCORRELATED}{temp\_correlated\_tuples}
%
%	\SetKwInOut{Input}{Input}
%	\SetKwInOut{Output}{output}
%	\KwData{Dataset $d$, Query answer $A$, constraints $C$}
%	\Output{\CORRELATED}
%	
%	\CORRELATED = \TEMPCORRELATED = $\emptyset$	\\	
%	{unvisited} = d \\
%	\While{!{\TEMPCORRELATED}.isEmpty}{
%		{\LHSSET = $A$.map(tuple $\rightarrow$ tuple(lhsIndex))}\\
%		{\TEMPCORRELATED = $unvisited$.filter(tuple $\rightarrow$ \LHSSET.contains(tuple(lhsIndex)))\\}
%		unvisited = unvisited - {\TEMPCORRELATED}
%		
%		{\RHSSET = $A$.map(tuple $\rightarrow$ tuple(rhsIndex))}\\
%		{\TEMPCORRELATED = $unvisisted$.filter(tuple $\rightarrow$ \LHSSET.contains(tuple(rhsIndex)))\\}
%		unvisited = unvisited - {\TEMPCORRELATED}
%	}
%	\caption{SP Query relaxation for FDs}
%	\label{query_relax}
%\end{algorithm}
\algrenewcommand\algorithmicindent{1em}%

%\lipsum[1-2]
{
	\setlength{\textfloatsep}{-10pt}% Remove \textfloatsep
%\begin{figure}
\begin{algorithm}[t]
	\caption{SP query result relaxation for FDs}
\begin{algorithmic}[1]
	\footnotesize
	\Require Dataset \textit{d}, Query answer \textit{A}, \textit{FD(lhs,rhs)}
	\State \textit{total\_extra} = $\emptyset$	
	\State \textit{extra = unvisited = d - A}
	\While {\textit{extra} $\neq$ $\emptyset$}
	\State $A_{lhs}$ = \textit{A}.map($x$ $\rightarrow$ $x_{lhs}$)
	\State $A_{rhs}$ = \textit{A}.map($x$ $\rightarrow$ $x_{rhs}$)
	
	\State \textit{extra} = \textit{unvisited}.filter($x$ $\rightarrow A_{lhs}$.contains($x_{lhs}$))
	\State \textit{unvisited} = \textit{unvisited} - \textit{extra}

	\State \textit{extra} = \textit{unvisited}.filter($x$ $\rightarrow A_{rhs}$.contains($x_{rhs}$))
	\State \textit{unvisited} = \textit{unvisited} - \textit{extra}
	\State \textit{total\_extra} = \textit{total\_extra} $\cup$ \textit{extra}
	\EndWhile
	\State \Return \textit{total\_extra}
\end{algorithmic}
\vspace{-0.5em}
\label{algo:query_relax}
%\vspace{-1.em}
\end{algorithm}
%\end{figure}

}

Algorithm~\ref{algo:query_relax} shows the general query result relaxation that uses transitive closure; it iteratively computes
the correlated tuples of the result, until 
no more correlated tuples are detected. 
Consider an FD $lhs$$\rightarrow$$rhs$, and $A_{lhs}$, $A_{rhs}$ being the set of left-hand-side (\textit{lhs}) and right-hand-side (\textit{rhs}) attribute values that appear in the result (lines 4,5). 
Algorithm \ref{algo:query_relax} traverses the data subset that does not belong to the relaxed result (unvisited) (lines 2,9) and enhances the result with each tuple $x$ for which \textit{\{$x_{lhs}$$\in$$A_{lhs}$\}} or \textit{\{$x_{rhs}$ $\in$$A_{rhs}$\}}
(lines 6-10).
%Thus, by computing the correlated tuples, $clean_{\sigma}$ determines the probabilistic candidate fixes of the accessed erroneous cells.
%by exploiting the correlation information among the tuples. 

The second step of $clean_\sigma$ is to detect errors and compute fixes given the relaxed result. Consider random variables \textit{LHS}, 
\textit{RHS} that represent the candidate \textit{lhs} and \textit{rhs} values of an erroneous tuple \textit{t}. \textit{LHS} contains the \textit{lhs} values of the tuples $t^\prime$
for which $t^\prime_{rhs}$=$t_{rhs}$, i.e., they have the same \textit{rhs} value. 
\textit{RHS} contains the \textit{rhs} values of the tuples $t^\prime$
for which $t^\prime_{lhs}$ =$t_{lhs}$.
Hence, by including all correlated values, future accesses to the cleaned tuples require no extra checks.
Candidates \textit{$c_{lhs}$$\in$LHS}, \textit{$c_{rhs}$$\in$RHS} have probabilities
P($c_{lhs}$$|$$t_{rhs}$), P($c_{rhs}$$|$$t_{lhs}$), respectively.
Thus, based on attribute dependencies, each tuple can have two instances, one with the candidate $c_{lhs}$, given the existing $t_{rhs}$
and one with the candidate $c_{rhs}$, given the existing $t_{lhs}$. As in our internal representation we use attribute-level uncertainty, we store
inside each candidate value the ID of the candidate pair it belongs.
%Thus, the candidate values for the $lhs$ and $rhs$ since they have equivalent values for a subset of the FD attributes. 

As Algorithm \ref{algo:query_relax} is iterative, we need to estimate the number of iterations required, as well as the relaxed result size, to accurately compute the fixes using the correlated tuples.

\vspace{-0.5em}
{\setlength{\parindent}{0em}
\begin{lemma}
\label{lemma:sp}
Algorithm~\ref{algo:query_relax} requires one iteration to enable accurate candidate fixes in the presence of SP queries with a filter on the $rhs$ of an FD.
\end{lemma}}
%\vspace{-1em}

{\setlength{\parindent}{0em}
\begin{proof}
	Consider a query with a filter restricting the \textit{rhs} over the range \textit{[a,b]}. The 
	correct result must include both the clean tuples with $rhs$ values in the range \textit{[a,b]},
	as well as the dirty tuples that are candidates to take values in \textit{[a,b]}.
	%Without the relaxation of the dirty answer and the computation of the probabilities, the query answer might contain 
	%tuples that are more likely to get a value outside the range $[a-b]$, and at the same time, it might ignore
	%tuples that could get a value in $[a-b]$ with high probability.
	%To prove that there are no missing tuples from the query answer we need to show that the answer relaxation mechanism
	%includes all the tuples that are candidates to take a value in the range $[a-b]$ for the $rhs$ attribute.
	Algorithm \ref{algo:query_relax} computes the tuples that have matching
	\textit{lhs} with the dirty result (line 6). We assume that, to exploit the dependency
	and be able to make a prediction, the dirty tuples 
	contain either a clean \textit{lhs}, or a clean \textit{rhs} \cite{scare}.
	The extra tuples with matching \textit{lhs} are the candidates to get \textit{rhs} in \textit{[a,b]}.
	Hence, enhancing the result with tuples having the same \textit{lhs} guarantees no missing tuples.
	%	Apart from proving that the algorithm includes the qualifying tuples, 
	We also show that 
	the included tuples contain all candidate values. 
	%In the following, we show that
	%the set of candidate values for each erroneous value of \textit{lhs} corresponds to the \textit{lhs} values of
	%the tuples which have the same value on the \textit{rhs}. Similarly, we show that
	%the candidate values for the \textit{rhs} of an erroneous tuple corresponds to the \textit{rhs} values of the tuples
	%with the same value on the \textit{lhs} attribute.
	Algorithm~\ref{algo:query_relax} covers the \textit{rhs} candidate values by computing the tuples with matching \textit{lhs}.
	Then, Algorithm~\ref{algo:query_relax} computes the \textit{lhs} values of all 
	tuples with matching \textit{rhs}. However, these tuples are already included in the enhanced-result as they satisfy the query;
	thus the algorithm terminates. 
	%Thus, in the presence of queries with conditions
	%on the \textit{rhs}, the resulting tuples contain all the candidate values.
\end{proof}}
\vspace{-0.3em}

\begin{table}[t]
	\begin{minipage}{.2\textwidth}
		%		\centering
		\footnotesize
		\begin{tabular}{|c|c|}
			\hline
			\textbf{Zip} & \textbf{City}\\
			\hline
			9001 & Los Angeles\\
			\hline
			9001  & San Francisco\\
			\hline
			9001  & Los Angeles\\
			\hline
			10001 & San Francisco\\
			\hline   
			10001 & New York\\
			\hline   
		\end{tabular}
		\subcaption{}
		\label{tab:dirty_cities}	
	\end{minipage}
	%	\qquad
	\begin{minipage}{.23\textwidth}
		\footnotesize
		\begin{tabular}{|c|c|}
			\hline
			\textbf{Zip} & \textbf{City}\\
			\hline
			\pbox{10cm}{9001} & \pbox{10cm}{Los Angeles, 67\%\\ San Francisco, 33\%} \\
			\hline
			%			\pbox{10cm}{9001, (a,100\%) (b,50\% )\\10001, (b,50\%)}  & \pbox{10cm}{Los Angeles, (a,67\%)\\ San Francisco, (a, 33\%) (b, 50\%)}\\
			\pbox{20cm}{9001}  & \pbox{20cm}{Los Angeles, 67\%\\ San Francisco, 33\%}\\ \hdashline
			\pbox{20cm}{9001 50\% \\ 10001 50\%}  & \pbox{20cm}{San Francisco}\\
			\hline
			\pbox{20cm}{9001}  & \pbox{20cm}{Los Angeles, 67\%\\ San Francisco, 33\%}\\
			\hline
			10001 & San Francisco\\
			\hline
			10001 & New York\\
			\hline
		\end{tabular}
		\subcaption{}
		\label{tab:cities_prob}	
	\end{minipage}
	\vspace{-0.6em}	
	\caption{Cities dataset: (a) Dirty version, (b) Partially clean version with candidate values.
		The dashed line denotes different candidate fixes for the erroneous tuples.}
	\label{tab:cities_dataset}
	\vspace{-1.9em}
\end{table}	

%\vspace{-1em}
{\setlength{\parindent}{0em}
\begin{example}
\label{ex:sp}
Consider the dataset of Table~\ref{tab:dirty_cities}, the FD $Zip$ $\rightarrow$ $City$,
and a query requesting the zip code of ``Los Angeles''.
\end{example}}
\vspace{-0.5em}

%\begin{verbatim}
%SELECT Zip, City
%FROM Cities
%WHERE City="Los Angeles"
%\end{verbatim}
%%\vspace{-0.2em}

%\noindent The dirty result contains the first and the third tuple of Table~\ref{tab:dirty_cities}.
%$clean_{\sigma}$, by following Algorithm \ref{algo:query_relax}, enhances the result with the tuples for which
%\textit{\{City$\neq$Los Angeles$\wedge$Zip=9001\}= \{9001, San Francisco\}}. Afterwards,  
%it adds the tuples of the set \textit{\{Zip$\neq$9001$\wedge$City=Los Angeles\}=$\emptyset$}.
%%However, based on the proof of Lemma\ref{lemma:sp}, this set is empty since the tuples with \textit{City=Los Angeles} already appear in the result.
%%The fourth tuple is excluded since it does not satisfy the query.
%Then, $clean_{\sigma}$ computes the probabilities \textit{P(City$|$Zip)} and \textit{P(Zip$|$City)} for the tuples of the
%updated result to calculate the probability of each candidate. For the first and third tuple, these probabilities are \textit{P(City=Los Angeles$|$Zip=9001)} and \textit{P(Zip=9001$|$City=Los Angeles)}. 
%For the second tuple there are two candidate
%pairs distinguished by a dashed line in the table:
%\textit{P(City$|$Zip=9001)= \{San Francisco 33\%,  Los Angeles 67\% \}}  and $P(Zip|City=San\  Francisco) = \{9001\: 50\%, 10001\: 50\% \}$. 
%The updated version of the dataset is shown in Table~\ref{tab:cities_prob}.

%%%%%%%%%%%%%%%%%%%%%%%%ADD%%%%%%%%%%%%%%%%%%%%%%%%%%%%%%%%%%
\noindent The dirty result consists of the first and the third tuple of Table~\ref{tab:dirty_cities}.
$clean_{\sigma}$, by following Algorithm \ref{algo:query_relax}, enhances the result with the tuples that have the same \textit{lhs} with the result, that is
the tuples for which
\textit{\{City$\neq$Los Angeles$\wedge$Zip=9001\}=\{9001, San Francisco\}}. Afterwards,  
it adds the tuples of the set \textit{\{Zip$\neq$9001$\wedge$ City=Los Angeles\}=$\emptyset$}, that is the ones that share a \textit{rhs} value with the result.
However, based on the proof of Lemma\ref{lemma:sp}, this set is empty since the tuples with \textit{City=Los Angeles} already appear in the result.
%The fourth tuple is excluded since it does not satisfy the query.
Then, $clean_{\sigma}$ computes the candidate fixes \textit{LHS} and \textit{RHS}, and their candidate probabilities \textit{P(City$|$Zip)} and \textit{P(Zip$|$City)} for the tuples of the
updated result. For the first and the third tuple, \textit{LHS} consists of the candidate values of the tuples $t^\prime$ that have $t^\prime_{rhs}=$\textit{Los Angeles}.
Similarly, the \textit{RHS} consists of the candidate values of the tuples that have $t^\prime_{lhs}=9001$, that is \textit{San Fransisco, Los Angeles}.
The corresponding probabilities of each value are given by the conditional probabilities \textit{P(City$|$Zip=9001)} and \textit{P(Zip$|$City=Los Angeles)}.
For the second tuple there are two candidate
pairs distinguished by a dashed line in the table for simplicity:
\textit{\{City$|$Zip=9001\}= \{San Francisco 33\%,  Los Angeles 67\%\}}  and \textit{\{Zip$|$City=San\  Francisco\} = \{9001\: 50\%, 10001\: 50\%\}}. 
The updated version of the dataset is shown in Table~\ref{tab:cities_prob}.

A filter over the $lhs$ requires multiple iterations in order to also include and fix
dirty tuples that qualify the query as well as their context.

%%%%%%%%%%%%%%%%%%%%%%%%%%%%%%%%%%ADD%%%%%%%%%%%%%%%%%%%%%%%%%%%%%%%%%%%%%%
{\setlength{\parindent}{0em} 
\vspace{-0.3em}
\begin{example}
	Consider the dataset of Table~\ref{tab:dirty_cities} and a query requesting the city name with zip code ``9001''.
\end{example}
\vspace{-0.5em}
}

%%%%%%%%%%%%%%%%%%ADD%%%%%%%%%%%%%%%%%%%%%%%%%%%
\noindent The dirty result comprises the first three tuples of Table~\ref{tab:cities_prob}. 
%%%%%%%%%%%%%%%%%%%%%%%%%%ADD%%%%%%%%%%%%%%%%%%%%%%%%%%
However, given the conflict between the tuples with zip code $10001$, the correct result contains
the four tuples shown in Table~\ref{tab:correct_lhs_answer}. The fourth tuple qualifies because
it has two worlds (\textit{\{\{90001 50\%, 10001 50\%\}, \{10001\}\}}), and the first one satisfies
the condition.
Thus, Algorithm~\ref{algo:query_relax} adds the tuple \textit{\{10001, San Francisco\}}
since it contains a $rhs$ value which appears in the result. Then, the next iteration adds the tuple 
\textit{\{10001, New York\}} since \textit{10001} belongs to the relaxed result. Thus,
using transitive closure, Algorithm \ref{algo:query_relax} determines the whole cluster of correlated entities.

\vspace{-0.4em}
{\setlength{\parindent}{0em}
\begin{lemma}
\label{lemma:iteration}
Consider a query with a filter on the \textit{lhs}, and a relaxed result $A_R$ with maximal size $|A_R|$ at iteration $i$.
Algorithm \ref{algo:query_relax} requires an extra iteration to compute the candidate values with probability $Pr(\geq 1) = 1-{\binom{\#vio}{0}\; \binom{n-\#vio}{|A_R|}}/ {\binom{n}{|A_R|}}$, where \textit{Pr} is the hypergeometric distribution,
$n$ the dataset size and the dataset has $\#vio$ violations.
\end{lemma}}
\vspace{-0.6em}

%\noindent The proof of Lemma \ref{lemma:iteration} can be found in \textcolor{red}{ADD}.
%%%%%%%%%%%%%%%%%%%%%%ADD%%%%%%%%%%%%%%%%%%%%%%%%%%%
{\setlength{\parindent}{0em}
\begin{proof}
	%To compute the upper bound in case of conditions over $lhs$, we estimate 
	%the number of iterations required to compute the total set of correlated tuples. 
	Algorithm \ref{algo:query_relax} terminates when the computed augmented result contains no more errors, that is
	there are no tuples with the same \textit{lhs} and different \textit{rhs}.
	Consider iteration $i$, where the relaxed answer $A_R$ has maximal result size $|A_R|$. The probability
	that $A_R$ contains at least one violation is equivalent to the complement of the probability of having no violations $Pr(0)$.
	Using the hypergeometric distribution, we estimate $Pr(0)$ over the subset $A_R$, given a total population of size $n$
	that contains $\#vio$ violations. Thus, $Pr(\geq 1) = 1-{\binom{\#vio}{0}\; \binom{n-\#vio}{|A_R|}}/ {\binom{n}{|A_R|}}$.
	%The number of iterations depends on the probability that the enhanced tuples are erroneous; 
	%a high number of violations increases the probability of having
	%a tuple of the answer interacting with another erroneous tuple. The probability
	%of accessing a tuple that violates a FD is:
	%$p_{vio} = \#violations/dataset\ size$. The probability $p_{vio}$ decreases while accessing and cleaning a bigger part of the dataset.
	%Thus, assume $c_{rhs}$ is the cardinality 
	%of \textit{rhs} in the query answer, and given that the selectivity of \textit{rhs} over the whole dataset is $s$, the  percentage of correlated tuples is:
	%$\mathcal{P}_c = c_{rhs} \ s \ p_{vio}$. 
	%At a high level, the percentage of extra values is equivalent to the probability that a tuple outside
	%the query result which has the same value on the \textit{rhs}, is erroneous. Thus, the probability depends on the distinct number
	%of \textit{rhs} that appear in the violation, the number of times that each value might appear over the whole dataset,
	%and the probability that the extra values are erroneous.
	%Therefore, the total set of correlated tuples is: $\mathcal{R}\,P_{c}$. 
\end{proof}
}
\vspace{-0.5em}

Thus, while cleaning the qualifying result, Algorithm \ref{algo:query_relax} might also detect and repair extra violations of
the correlated tuples of the result.

%%%%%%%%%%%%%%%%%%%%%%%%%%%%%%%ADD%%%%%%%%%%%%%%%%%%%%%%
\begin{table}[t]
	\centering	
	\footnotesize
	\begin{tabular}{|c|c|}
		\hline
		\textbf{Zip} & \textbf{City}\\
		\hline
		\pbox{10cm}{9001} & \pbox{10cm}{Los Angeles, 67\% San Francisco, 33\%} \\
		\hline
		%			\pbox{10cm}{9001, (a,100\%) (b,50\% )\\10001, (b,50\%)}  & \pbox{10cm}{Los Angeles, (a,67\%)\\ San Francisco, (a, 33\%) (b, 50\%)}\\
		\pbox{20cm}{9001}  & \pbox{20cm}{Los Angeles, 67\% San Francisco, 33\%}\\ \hdashline
		\pbox{20cm}{9001 50\%,  10001 50\%}  & \pbox{20cm}{San Francisco}\\
		\hline
		\pbox{20cm}{9001}  & \pbox{20cm}{Los Angeles, 67\% San Francisco, 33\%}\\
		\hline
		10001 & \pbox{20cm}{San Francisco, 50\% New York, 50\%}\\ \hdashline
		\pbox{20cm}{9001 50\%, 10001 50\%}  & \pbox{20cm}{San Francisco}\\
		\hline		
	\end{tabular}
	%		\subcaption{Accurate version of the query answer.}
	%		\label{tab:correct_answer}	
	%	\end{minipage}
	\caption{Correct query result given condition on the \textit{lhs}. The query result becomes accurate after traversing the dataset again to fetch more correlated entities.}
	\label{tab:correct_lhs_answer}
	\vspace{-2em}
\end{table} 

%%%%%%%%%%%%%%%%%%%%%%%ADD%%%%%%%%%%%%%%%%%%%%%%%
\vspace{-0.5em} 
{\setlength{\parindent}{0em}
	\begin{lemma}
		\label{lemma:upper_bound}
		%The upper bound of the query answer relaxation is as follows.
		Let $\mathcal{A}$ be the set of attributes that appear in the FDs.
		Let $c_i$ be the cardinality (number of distinct values) of each attribute $A_i$$\in$$\mathcal{A}$ in the query result, 
		and $D_i$, $Dq_i$ the frequency distributions of each $A_i$ over the dataset and the query result respectively.
		The upper bound of the relaxed result size in each iteration is $\mathcal{R} =\sum_{i=1}^{|\mathcal{A}|} (\sum_{j=1}^{j=c_i} D_{ij} - \sum_{j=1}^{j=c_i} Dq_{ij})$.
	\end{lemma} }
	\vspace{-0.5em}
	
	%\noindent We provide the proof of Lemma \ref{lemma:upper_bound} in \textcolor{red}{\cite{??}}.
	%%%%%%%%%%%%%%%%%%%%%%%%%%%%%%%%%%ADD%%%%%%%%%%%%%%%%%%%%%%%%%%%%%%%%%%%
	\vspace{-0.3em}
	{\setlength{\parindent}{0em}
		\begin{proof}
			Given that the $c_i$ values of the result follow a distribution $Dq_i$, then 
			the total frequency of these values over the dataset is $\sum_{j=1}^{j=c_i} D_{ij}$.
			The upper bound corresponds to the worst case scenario where there is no overlap between the sets of 
			correlated tuples stemming from each attribute $A_i$.
			In the worst case, the number of extra tuples that the relaxation adds to the result set corresponds to the 
			number of tuples sharing the same value for each attribute of the result. Therefore, in total, the number of tuples is: $\mathcal{R} =\sum_{i=1}^{|\mathcal{A}|} (\sum_{j=1}^{j=c_i} D_{ij} - \sum_{j=1}^{j=c_i} Dq_{ij})$.
			%This value corresponds to an upper bound because it assumes no overlap between the sets of tuples corresponding to each attribute $A_i$.
		\end{proof}}
		\vspace{-0.5em}
		
		%%%%%%%%%%%%%%%%%%%%%%%%%%%%%%%%%ADD%%%%%%%%%%%%%%%%%%%
		\noindent Thus, in the case of queries restricting the $rhs$, the upper bound is equivalent to $\mathcal{R}$.

%The intuition behind the computation of the candidate values that might fit in each erroneous cell follows the principle of minimal repairs \cite{data_quality}.
%The principle of minimal repairs states that a repair of a dirty dataset should minimally differ from it. 
%Therefore, we assume that at least a subset of the
%attribute values that participate in a violation of a rule must be correct. This principle justifies the computation of the conditional probabilities 
%between the $lhs$ and $rhs$ attributes since it is driven by their correlation.

%The main benefit of this approach is that when executing each query, it enhances the query answer with the set of tuples that are required for cleaning the 
%query answer. 

	\noindent\textbf{Relaxation benefit:} The extra tuples contain the pruned domain of values that a system, or a user needs to infer the correct value of
	an erroneous cell \cite{holoclean}.
	Specifically,
	a query result contains a set of tuples with a restricted set of values for the attributes of 
	the constraints.
	%specific characteristic, e.g., tuples with an equality or
	%range filter on a specific attribute, 
	The cleaning process can exploit this characteristic and extract all the correlated tuples of the result, instead of computing the candidate fixes separately
	for each violated tuple.
	%%%%%%%%%%%%%%%%%%%%%ADD%%%%%%%%%%%%%%%%%5
	Thus, instead of traversing the whole dataset for each erroneous value to compute the candidate fixes, relaxation iterates over the correlated tuples. 

%In the following we formalize the query relaxation technique depending on the attributes that an SP query might access. We assume the following syntax:

%\begin{verbatim}
%SELECT PLIST
%FROM FROMLIST
%WHERE filter WHERELIST
%\end{verbatim}
%More specifically, after having grouped the elements of the dataset 

%The first three tuples violate the functional dependency and th

\subsection{Cleaning SP query results given a DC}
\label{subsec:relax_dcs}
\vspace{-0.2em}

We present the case of more general rules with arbitrary predicates.
$clean_\sigma$ first computes the correlated tuples that, in the case of DCs, involve the conflicting tuples with the query result.
Detecting the correlated tuples requires a self theta-join. 
%a cartesian product would be required, followed
%by a filter that checks if the condition (e.g., inequality) is satisfied. 
We adopt an optimized parallel theta-join 
approach \cite{mr_joins} that maps the cartesian product to 
a matrix that it partitions into \textit{p} uniform partitions.
Using the matrix, we check arbitrary predicates between any pair of attributes.
In our analysis, we focus on the more realistic case that involves conditions over the same attribute~\cite{bigdansing}.
We compute incrementally the theta-join, by partitioning and checking the matrix subset
that affects the result; the matrix subset involves the query result and the unseen part of the dataset. We
also prune the redundant symmetric parts of the matrix.

\begin{figure}[t]
	\centering
	\begin{minipage}{.19\textwidth}
		\includegraphics[width=\textwidth]{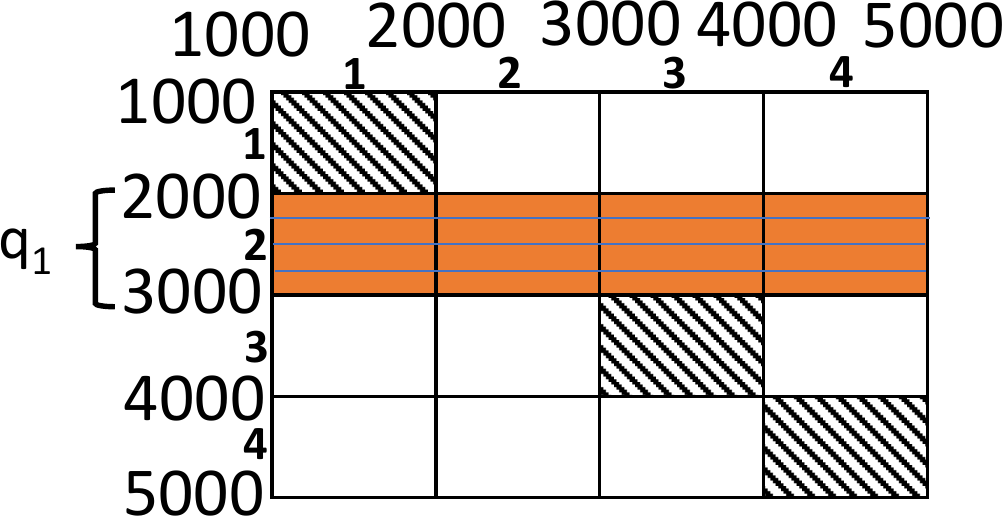}		
		\vspace{-1.8em}		
		\caption{Cleaning $q_1$}
		\label{fig:dc_q1}	
	\end{minipage}
	\hspace{0.7em}
	\begin{minipage}{.18\textwidth}
		\includegraphics[width=\textwidth]{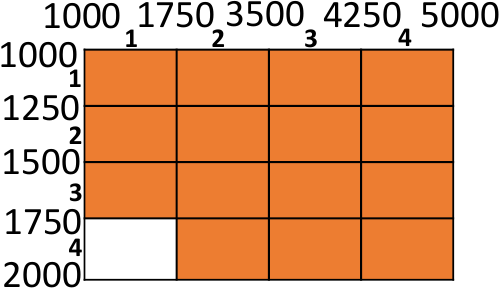}				
		\vspace{-1.8em}		
		\caption{Cleaning $q_2$}
		\label{fig:dc_q2}	
	\end{minipage}			
	%		\vspace{-1em}		
	%	\caption{Incremental cleaning in the presence of general DCs.}	
	\vspace{-1.5em}		
\end{figure}

Partial theta-join operates at a finer granularity, hence it can prune a) matrix partitions and b) pairs within a partition, which are not candidates for conflicts.
%Theta-join prunes partitions for which the boundaries do not qualify the condition \cite{mr_joins}
First, by partitioning a subset of the matrix, partitions have boundary ranges smaller than the more general boundaries of the original matrix partitions; 
there might exist sub-partitions whose boundaries do not qualify the condition, even though the general partition qualifies.
Second, partial theta-join prunes non-qualifying intra-partition pairs; within a partition, it restricts the candidate pairs to be checked.

\vspace{-0.7em}
{\setlength{\parindent}{0em}
	\begin{example}
		\label{ex:dc_cartesian}
		Consider a dataset with salary, tax values and rule $\phi$:$\forall$$t_1$,$t_2$:$\urcorner$($t_1$.salary$<$$t_2$.salary$\wedge$$t_1$.tax$>$$t_2$.tax).
		Fig.~\ref{fig:dc_q1} shows an example cartesian product matrix based on rule $\phi$. Consider two queries requesting salary ranges $[2000-3000]$ and $[1000-2000]$
		respectively. 
	\end{example}
}
\vspace{-0.5em}

\noindent To clean the first query result, theta-join checks for violations in the orange area of Fig. \ref{fig:dc_q1} which it
divides into $p$ partitions. Then, for the second query it constructs a matrix with vertical range (1000-5000)$\smallsetminus$(2000-3000) since the subset (2000-3000)
has been already checked, thus it excludes it from the comparisons. Given a smaller range, the boundaries will be the ones of Fig. \ref{fig:dc_q2}. Theta-join can then filter out non-qualifying partitions, such as partition (4,1). 
It also applies intra-partition filtering to exclude non-qualifying pairs. For example, given partition (3,1) with horizontal and vertical 
ranges (1500, 1750) and (1000, 1750) respectively, since we are interested in checking the $<$ condition, the vertical range is transformed into (1500, 1750) since the part (1000, 1500) will not produce candidate violations.
%By dividing subsets of the matrix into partitions, the algorithm
%performs fewer checks since it can further filter out the non-qualifying parts of each partition. For example given the partition with ranges (275, 300) and (200, 300), the subset 

The second step involves computing the candidate fixes. For DCs we use the holistic data 
cleaning approach~\cite{holistic_dc} to calculate the possible conditions that the dirty cells must satisfy. Specifically, given a rule with
inequality predicates, $clean_{\sigma}$ replaces the errors with the candidate ranges that satisfy the constraints. 
Then, similarly to FDs, the probability of each candidate is frequency-based,
given the total number of fixes.

More formally, given a DC \textit{$\forall$$t_i$,$t_j$$\neg$($t_i.v_1$>$t_j.v_2$)} and two tuples $t_1$,$t_2$ for which $t_1.v_1$>$t_2.v_2$, then a candidate fix 
of the violation needs to enforce the constraint; \textit{$t_1.v_1$=\{{$t_1.v_1$ or <$t_2.v_2$\}}}, \textit{$t_2.v_2$=\{{$t_2.v_2$ or >$t_1.v_1$\}}}.  
Thus, each attribute value will either maintain its original value, or will obtain a value satisfying the range.
In the case of DCs that contain more atoms, we map the dirty formula involving the conditions of the conflicting tuples to a SAT formula \cite{nadeef}, where a subset of atoms must become false (invert their condition) in order to satisfy the formula.
Thus, a possible violation fix requires updating the appropriate attribute values in order to invert the condition of the subset of atoms
that cause the violation. 
Thus, to include all the possible combinations of violated atoms, we produce all possible candidate attribute combinations.
Then, a SAT solver \cite{sat} can decide on which atoms must remain true or need to invert their conditions (become false)
in order to satisfy the whole DC formula.

The probabilities of each candidate fix are based on the frequency that each of the candidate ranges appears.
We provide frequency-based probabilities to collect all possible fixes for a specific value, accompanied by their weight.  
Then, after having computed the candidate fixes of the data subset that affects the cleanliness of a value $v_i$, an 
inference algorithm can repair the dirty values. Future work considers updating the probabilities after accessing
more data, thereby incrementally inferring the correct value.

\vspace{-0.7em}
{\setlength{\parindent}{0em}
	\begin{example}
		\label{ex:dc}
		Consider a dataset with \{salary,tax,age\} values $t_1$: \{sal:1000,tax:0.1,age:31\}, $t_2$:\{sal:3000,tax:0.2,age:32\}, $t_3$:\{sal:2000, tax:0.3,age:43\}
		and $\phi$:$\forall$$t_1$,$t_2$:$\urcorner$($t_1$.salary$<$$t_2$.salary$\wedge$$t_1$.tax$>$$t_2$.tax).
		%Then, given
		%a SP query requesting for employees with tax 0.1\%, then, the answer would contain only the first tuple, even though $t_3$
		%might contain the value $0.1$ due to the violation with tuple $t2$ ($t3.tax$ must be $<0.2$).
\end{example}}
\vspace{-0.8em}

\noindent Tuples $t_2$,$t_3$
	violate $\phi$, thus the candidate fixes for $t_2$ are \{($<$2000 50\%,3000 50\%),0.2,32\},\{3000,(0.2 50\%,$>$0.3 50\%), 32\},
	that is, $t_2$ must either take a salary less than 2000 or have tax greater than 0.3. The probabilities
	stem from the fact that there are two possible fixes.
	%%%%%%%%%%%%%%%%%%%%%%%%%%%ADD%%%%%%%%%%%%%%%%%
	Given a DC with more than two atoms, the candidate values contain the conditional probabilities
	of all possible subsets of atoms. For example, given $\phi_2$:$\forall$$t_1$,$t_2$:$\urcorner$($t_1$.salary$<$$t_2$.salary$\wedge$$t_1$.age$<$$t_2$.age$\wedge$$t_1$.tax $>$$t_2$.tax)
	which requires that both the salary and the age of the employee define her tax rate, then apart from the aforementioned candidates, we need to
	include the respective fix of the age field (\{3000,0.2,(32 50\%,$>$43 50\%)\} followed by the pairwise combinations of all three candidate fixes.

\algrenewcommand\algorithmicindent{1em}%

{
	\setlength{\textfloatsep}{-10pt}% Remove \textfloatsep
%\begin{figure}[t]
\begin{algorithm}[t]
	\caption{Query-driven cleaning DC violations}
\begin{algorithmic}[1]
	\footnotesize
	\Function{\textit{Estimate\_Errors}}{\textit{data d, partitions p, rules r}}
%	\Require Dataset $d$, partitions $p$, constraints $C$ 	
	\State \textit{ranges = split(d,p)}
	\For{\textit{$r_1$ in ranges}}
	  \For{\textit{$r_2$ in ranges}}
		   	\If{\textit{overlap($r_1$, $r_2$)}}
			   	\State \textit{range\_vio($r_1$) = count\_overlap($r_1, r_2, r$)}
		   	\EndIf
	   	\EndFor   
	\EndFor
	\Return{range\_vio}
	\EndFunction
\end{algorithmic}
\vspace{-0.7em}
\begin{algorithmic}[1]
	\footnotesize	
	\Require{\textit{queries queries, data d, partitions p, rules r, threshold th}}
	\State \textit{range\_vio = Estimate\_Errors(d,p,r)}
	\For{\textit{query in queries}}
		\State \textit{$q_a$ = execute query}
		\State \textit{range = find range of $q_a$ in range\_vio}
		\State $errors$ = \textit{\{for (i $\neq$ range) yield range\_vio(i)\}.sum}
		\State \textit{accuracy = errors/($|q_a|$+errors)}
		\State $support=(1+2+...+\sqrt{p})-unchecked\_p/(1+2+...+\sqrt{p})$
		\If{\textit{accuracy $>$ th}}
			\textit{full cleaning}	
		\Else{}
			\State \textit{partial cleaning}					
		\EndIf		
%		\State \Return $total\_extra$		
	\EndFor
\end{algorithmic}

\vspace{-0.7em}

\label{algo:dc}
\vspace{-0.5em}
\end{algorithm}
%\end{figure}
}

%\vspace{-1em}
\noindent\textbf{Accuracy:} %The accuracy of the query answer in the case of DCs depends on the number of violations of the dataset.
DC violations affect result quality since a dirty value might get a candidate fix that satisfies the query. 
%If we consider the dataset of Example \ref{ex:dc} and
%a SP query requesting for employees with tax 0.1\%, the result misses tuple $t_3$
%which might also obtain value $0.1$ due to the violation with tuple $t_2$ ($t_3.tax$ might be $<0.2$).
To compute result accuracy, $clean_{\sigma}$ estimates the theta-join selectivity using the \textit{Estimate\_Errors} function of Algorithm~\ref{algo:dc}.
The function takes as input the matrix partitions and calculates the overlap of the partition boundaries,
that is the number of conflicts between them \cite{mr_joins,inequality_joins}. 
For example, consider ranges 3 and 4 of Fig. \ref{fig:dc_q1}, with salary boundaries (3000,4000), (4000,5000) and tax boundaries 
	(0.3,0.4), (0.25,0.5) respectively. The violations lie in the overlap of tax values, that is, (0.25,0.4).
Thus, given
query answer $q_a$ (line 3), we identify the ranges with which $q_a$ overlaps (e.g., $q_1$ result overlaps with range 2), and obtain the total estimated errors for
these ranges.
Assuming inequality conditions, the erroneous partitions that affect the result are the ones with both a \textit{row} and a \textit{column} smaller or larger than the \textit{row/column} of the current range. Otherwise, the erroneous partitions will contain  either smaller or larger candidate value ranges than the result range.
Then, we compute if the estimated accuracy (line 6) exceeds the given threshold (input by the user)
and decide to fully or partially clean the data. 
The range overlap of \textit{Estimate\_Errors} function is only applicable over the non-diagonal partitions, since for the diagonal partitions (pattern filled boxes) the ranges are equivalent,
thus we also provide
the support, that is, the percentage of checked diagonal partitions. 
%Specifically, the statistics
%provide the pairs belonging to the blocks of the matrix excluding the blocks of the diagonal. 
The support
is defined as the total partitions checked {\footnotesize $(1+2+...+\sqrt{p})$}, which are the upper/lower diagonal partitions, minus the blocks of the diagonal ({\footnotesize $\sqrt{p}$}) in the first iteration and becomes smaller
depending on the accessed data (line 7).
Accuracy also increases while accessing and cleaning more entities.

%More formally, the accuracy of the $i^{th}$ query in the case of DCs can be defined as the number of violations that 
%have been cleaned divided by the estimated number of violations: $\sum_{j=1}^{j=i-1} \epsilon_j/\#total\ violations$.
%\TODO{say that we switch between full/incremental in cost model}

\vspace{-0.3em}
\subsection{Cleaning SP results given multiple DCs}
\label{subsec:multiple_dcs}
\vspace{-0.3em}

%\noindent\textbf{Multiple rules: } 
In the case of multiple rules, an erroneous cell might trigger violations of many of these rules.
Thus, the probability of each fix must combine the probabilities
that stem from all the rules affecting the erroneous cell. 
%However, for performance reasons, and when the accuracy guarantees permit, we
%can partially clean a cell by considering a subset of the constraints that affect the values of the cell. The accuracy of partially cleaning
%the constraints is explained in Section \ref{sec:accuracy}.
%In the presence of multiple constraints, the selection of the enhanced set of tuples, as well as the computation of the probabilities of the candidate values for each erroneous cell need to
%take into consideration the overlap that the set of functional dependencies might have. The types of overlap between two functional dependencies $\phi_1$
%and $\phi_2$, might be: $\phi_1$ and $\phi_2$ might overlap on their $lhs$ attributes, on their $rhs$ attributes,
%or the $lhs$ of one constraint might overlap with the $rhs$ of the other; 
%$\phi_1: X \rightarrow Y$, and $\phi_2: X \rightarrow Z$,
%$\phi_1: X \rightarrow Y$, and $\phi_2: Y \rightarrow Z$,
%$\phi_1: X \rightarrow Y$, and $\phi_2: Z \rightarrow Y$.
%The overlap between the constraints can be generalized for the case of multiple constraints.
For the dirty cells that belong to the overlapping attributes of multiple rules, we compute the  
candidate values in parallel then merge the resulting fixes.
We also maintain provenance information for each dirty cell; when many rules
exist, we execute them over the original data then merge with the already computed probabilities.
Finally, to prune unnecessary error checks, \system maintains information about the already checked tuples by each rule.

To merge the probabilities, we compute the overlap of the violating groups. 
We also use union to merge the candidate values of the overlapping cells,
and adjust the probabilities to reflect the union of the sets.
Given rules \textit{$\phi_1$:Y$\rightarrow$X}, \textit{$\phi_2$:Z$\rightarrow$X},
we assign $P(X|(Y \cup  Z))$ to the X values.
Thus, given \textit{zip$\rightarrow$state and city$\rightarrow$state} and two versions of a tuple with \textit{P(CA$|$9001)} and \textit{P(CA$|$LA)}, respectively, one for each rule,
the probability will be updated to \textit{P(CA$|$9001$\cup$LA)}, to match the tuples that have either zip \textit{9001} or city \textit{LA}.

%For example, assume the rules $\phi_1: X \rightarrow Y$, and $\phi_2: Z \rightarrow Y$, and an erroneous cell $\in Y$. Given that each candidate value $i$ of the cell
%appears $v_{i}$ times over a set of $n_{ij}$ values that either share the same $X$ or $Z$ value, and $j$ is the number of functional dependencies that the cell violates, 
%then we use the Bayes rule mentioned above to reflect the union of the rules. Therefore, the final set of probabilistic values for that cell
%will be:
%
%\[ c_i = \{\dfrac{v_{ij}}{\sum_{1}^{j} n_{ij}} | \forall j \in set\ of\ FDs \} \]

\vspace{-0.7em}
{\setlength{\parindent}{0em}
\begin{lemma}
	In the presence of multiple constraints, the order of computing the candidate values of the erroneous cells obeys the commutative property.
\end{lemma}}
%\vspace{-1.5em}

{\setlength{\parindent}{0em}
\begin{proof}
	Consider rules $\phi_1$ and $\phi_2$, which both involve attribute $X$, and a dirty tuple $e$ which
violates both rules. The probabilistic fix of
cell $e_x$$\in$$X$ of tuple $e$ based on both rules is the same regardless of the order that we check the rules.
Consider merging order $\phi_1$ followed by $\phi_2$. Based on $\phi_1$, $e_x$ becomes:
$e_x$=[($a_1$,$T_1$),($a_2$,$T_2$),...,($a_k$,$T_k$)], where $a_1...a_k$ are the candidate values of $e_x$, 
and each $T_i$ comprises the conflicting tuples due to which we assign value $a_i$. For example, for FDs,
$T_i$ involves the tuples with the same $lhs$ and different $rhs$.
Then, $\phi_2$ produces the corresponding set $e_{x\prime}$. The end result
contains the merge of $e_x, e_x\prime$, which involves pairs $(a_i, T_{mi})$, where $T_{m_i}$ is the union of $T_i$ and $T_{i\prime}$. Thus, since the
union is commutative, the result is independent of the rule order. 
\end{proof}}

\subsection{Cleaning Join results}

In the case of join queries, the cleaning operator needs to examine how the existence of
errors in each individual table affects the query result. 
The operator is defined as follows.

\vspace{-0.5em}
{\setlength{\parindent}{0em}
	\begin{definition}
		\label{def:join}
		$clean_{\bowtie}$ is a cleaning operator which cleans a join result. $clean_{\bowtie}$ (a) extracts
		the qualifying parts of the join tables, (b)
		cleans each part and updates each relation separately, (c) updates the result,
		and (d) re-checks for errors.
	\end{definition}}
\vspace{-0.5em}
%In addition, apart from cleaning each individual table based on the constraint violations, we need to take into consideration whether the violations 
%involve the join attributes; in the case of an overlap of the join attributes with the rule attributes, a violation might result in having missing tuples from the join result. 
%In the following, we describe the process of cleaning the answers of join queries, and explain how the cleaning operation might affect the query result.
\noindent Consider a join between $R$ and $S$.
To clean the dirty result, $clean_{\bowtie}$ extracts and cleans the corresponding qualifying parts of $R$ and $S$. 
To extract the qualifying parts of $R$ and $S$, $clean_{\bowtie}$ keeps provenance information~\cite{data_lineage} which allows to obtain the
entities of each table from the join result, as well as update the join result after cleaning the tables.
Thus, using lineage, after cleaning both tables, $clean_{\bowtie}$ recomputes the join 
to check whether the extra tuples of each relation produce new pairs. 
In the case where the cleaning task transforms the join key into a probabilistic attribute, the join becomes a probabilistic
join. % and outputs a pair if there is an overlap between the set of join keys of the tables.
In the following, we show that the updated join accesses the already clean tuples.
%For example, an extra tuple stemming from the enhanced set of relation $R$ might match with a tuple of relation $S$. 
%Therefore, the process of join and clean is iterative until the cleaning task does not produce any new value. In Section \ref{sec:cost_model} we show that 
%the maximum number of iterations required is two.

\begin{table}[t]
	\begin{minipage}{.22\textwidth}
		\centering	
		\footnotesize
		\begin{tabular}{c|c|c|}
			\cline{2-3}
			&\textbf{Zip} & \textbf{City}\\
			\cline{2-3}
			t1 & 9001 & Los Angeles\\
			\cline{2-3}
			t2 & 9001  & San Francisco\\
			\cline{2-3}	
			t3 & 10001 & San Francisco\\
			\cline{2-3}
		\end{tabular}
		\subcaption{Cities dataset.}			
		\label{tab:R_rel}
	\end{minipage}
	%	\quad
	\begin{minipage}{.22\textwidth}
		\centering
		\footnotesize
		\begin{tabular}{|c|c|c|}
			\hline
			\textbf{Zip} & \textbf{Name} & \textbf{Phone}\\
			\hline
			9001 & Peter & 23456 \\
			\hline
			10001 & Mary & 12345 \\
			\hline
			10002 & Jon & 12345 \\
			\hline
		\end{tabular}
		\subcaption{Employee dataset.}
		\label{tab:S_rel}	
	\end{minipage}
	\begin{minipage}{0.4\textwidth}
		\centering
		\footnotesize
		\begin{tabular}{|c|c|}
			\hline
			\textbf{Zip} & \textbf{Name}\\
			\hline
			9001 & Peter \\
			\hline
			%			9001 & Peter \\			
			%			\hline		
		\end{tabular}
		\subcaption{Dirty version of the join result.}
		\label{tab:join_dirty}	
	\end{minipage}
	\quad		
	\begin{minipage}{0.15\textwidth}
		%		\centering		
		\footnotesize
		\begin{tabular}{|c|}
			\hline
			\textbf{Zip}\\
			\hline
			\pbox{20cm}{9001}\\
			\hline
			\pbox{20cm}{9001, 50\%\\10001, 50\%}\\
			\hline									
		\end{tabular}
		\subcaption{Relaxed result of Select Operator over Cities.}
		\label{tab:join_it1}	
	\end{minipage}				
	\begin{minipage}{0.3\textwidth}
		\centering
		\footnotesize
		\begin{tabular}{|c|c|c|}
			\hline
			\textbf{C.Zip} & \textbf{E.Zip} & \textbf{Name}\\
			\hline
			\pbox{20cm}{9001} & 9001 & Peter \\
			\hline
			\pbox{20cm}{9001, 50\%\\10001, 50\%} & 9001 & Peter \\
			\hline
			\pbox{20cm}{9001, 50\%\\10001, 50\%} & \pbox{20cm}{10001, 50\%\\10002, 50\%} & Mary \\
			\hline			
			\pbox{20cm}{9001, 50\%\\10001, 50\%} &\pbox{20cm}{10001, 50\%\\10002, 50\%} & Jon \\
			\hline									
		\end{tabular}
		\subcaption{Clean join result.}
		\label{tab:join_it2}	
	\end{minipage}	
	\caption{Join operation over two tables that involve violations on the join key. }		
	\label{tab:join_ex}
	\vspace{-1.7em}
\end{table}

\vspace{-0.6em}
{\setlength{\parindent}{0em}
\begin{lemma}
	The updated join result stemming from the cleaned qualifying table parts, requires no extra violation checks. 
\end{lemma}}
\vspace{-1.1em}

%%%%%%%%%%%%%%%%%%%%%%%%%%ADD%%%%%%%%%%%%%%%%%%%%%%%%%%%%%%%%%%
{\setlength{\parindent}{0em}
\begin{proof}
	To prove the correctness of the result, we examine the possible correlation cases between the
	query and the rules. 
	%For simplicity, we refer to the attributes that participate in the constraints as erroneous attributes. 
	We assume that the result of any erroneous underlying operator in the plan has been cleaned.
	The possible scenarios depend on whether the join key appears in a constraint.
	When the join key is clean, the new tuples that relaxation produces
	will not qualify the join because they will contain a non-qualifying join key.
	If the join key attribute appears in a rule, then 
	consider a join of \textit{R} and \textit{S} which both involve errors on the join key.
	Then, $clean_{\bowtie}$ might add new tuples to both relations. However, the extra tuples of 
	$R$ will match with tuples of $S$ which already 
	exist in the result since they have to belong to the intersection of the join keys of $R$ and $S$. 
	Thus, no extra violations will exist.
	%since the added tuples have already been examined and cleaned. 
	The same case holds for the relaxed result of $S$.
\end{proof}}

%{\setlength{\parindent}{0em}
%\begin{proof}
%Assuming that the result of any erroneous underlying operator in the plan has been cleaned,
%either the extra tuples will join with existing cleaned tuples, or they will not qualify
%the join condition. The detailed proof can be found in \textcolor{red}{ADD}
%\end{proof}}

%%%%%%%%%%%%%%%%%%%%%%%%%%%%%%%%%%ADD%%%%%%%%%%%%%%%%%%%%%%%%%%%%%%%%%%%%%%%%%%%%%%%%%%%%%
\vspace{-1.1em}
{\setlength{\parindent}{0em}
\begin{example}
	Consider tables $Cities$ ($C$) and $Employee$ ($E$) shown in Tables \ref{tab:R_rel}, \ref{tab:S_rel}, rules \textit{$\phi_1$:Zip$\rightarrow$ City}, 
	\textit{$\phi_2$: Phone$\rightarrow$Zip},
	and a query requesting the name, and zip code from both Cities and Employee, for the city of ``Los Angeles'':
\end{example}}
\vspace{-0.5em}
%\begin{minted}[escapeinside=||,fontsize=\footnotesize]{mysql}
%SELECT C.Zip,E.Zip,E.Name
%FROM Cities C, Employee E
%WHERE C.Zip = E.Zip AND C.City = "Los Angeles"
%\end{minted}

\noindent The dirty version of the result is shown in Table \ref{tab:join_dirty}.
Similarly to Example \ref {ex:sp}, $clean_{\sigma}$ enhances the result with tuple $t_2$ as it belongs to the set 
\textit{\{City$\neq$Los\ Angeles$|$Zip = 9001\}}. Then, after repairing the detected errors of $\phi_1$, tuple $t_2$ of relation $C$ has
candidate values for the $Zip$  $\{9001\ 50\%, 100001\ 50\%\}$. 
The result of $clean_{\sigma}$ is shown in Table~\ref{tab:join_it1}.
Then, the evaluation of the join matches the filtered set 
of $C$ with all tuples of $E$ and $clean_{\bowtie}$ triggers the violation between tuples $t_2, t_3$ of $E$.
Thus, $clean_{\bowtie}$ fixes the corresponding part and updates the result.
The final version
of the result is shown in Table \ref{tab:join_it2}.

%The following sections present how we integrate the aforementioned cleaning operators inside the query plan,
%and the accuracy of the resulting query answers.

%\noindent\textbf{Summary:} Regardless of the type of the query, we need to enhance the query answer with extra tuples from the dataset in order to allow the detection and repairing of errors.
%The extra tuples comprise the tuples which are similar to the tuples belonging to the query answer.
%Future work includes examining query relaxation approaches for other types of operations, such as the general case of denial constraints \cite{data_quality_management}, which
%include arbitrary first order logic rules.

%In the following we analyze the maximum number of iterations required to produce a correct answer depending on the query plan that contains
%a join operator. More specifically, we distinguish among the cases where there is a filtering condition before the join operator which
%also affect the attributes of the functional dependency.
%\vspace{-1.5em}
\section{Cleaning-aware \\ Query Optimization \& Planning}
\label{sec:loglevel_optimizations}

In this section, we present how we inject cleaning operators at the logical level, and show a set of optimizations that enable 
an optimal placement of cleaning operators in the query plan.
We support queries with Select, Join and Group-by clauses.  %\TODO{add query template}
%%%%%%%%%%%%%%%%%%%%%%REMOVE%%%%%%%%%%%%%%%%%%%%%%%%%%%%%%%%%%%%%%%%%%%%%%%
%The template of the supported queries can be found in \textcolor{red}{\cite{??}}
%%%%%%%%%%%%%%%%%%%%%%%%%%ADD%%%%%%%%%%%%%%%%%%%%%%%%%%%%%%%%%%%%%%%%%%%%
The template of the supported queries is the following:
%\vspace{-0.5em}
%\begin{minted}[escapeinside=||,fontsize=\footnotesize]{sql}
{
	\footnotesize
\begin{verbatim}
SELECT <SELECTLIST>
FROM <table name> [,(<table name>)]
[WHERE <col><op><val> [(AND/OR <col><op><val>)]]
[GROUP BY CLAUSE]
\end{verbatim}
}
%\end{minted}
\vspace{-0.5em}
%%%%%%%%%%%%%%%%%%%%%%%ADD%%%%%%%%%%%%%%%%%%%%%%%%%%%
\noindent [] and () indicate optional and repeated elements. $op$ takes values $=,\neq, <, \leq, >, \geq$.
In the case of joins, we assume equi-joins.
We focus on flat queries to stress the overhead of the cleaning operators over the corresponding query operators.

\subsection{Cleaning operators in the query plan}
\label{subsec:cleaning_ops}

The logical planner detects the query operator attributes that appear in a rule and injects the appropriate
cleaning operators in the query plan. 
The planner pushes cleaning operators down, closer to the data, to avoid propagating errors in the plan. 
Deferring the execution of a cleaning task causes (a) redundant cleaning
to detect errors that have propagated from the underlying query operators, and/or
(b) recomputing the underlying query operators that are affected by the errors.
For example, consider relations $R$ and $S$, and a query with a select condition over attribute $R.a$ that participates in a rule, 
followed by a join $R.a \bowtie S.a$ where $S.a$ participates in a rule of $S$. Executing cleaning after the
join might alter the qualifying part of $R$ by adding an extra probabilistic tuple. If the extra tuple
matches with an unseen tuple of $S$, it will update the join result. Then, the operator needs to re-check for errors over the extra accessed tuples of $S$,
which induces a redundant overhead for query execution.
Hence, placing cleaning operators early in the plan avoids extra effort to fix propagated errors.

\begin{figure}[t]
	\centering
	\includegraphics[width=0.23\textwidth]{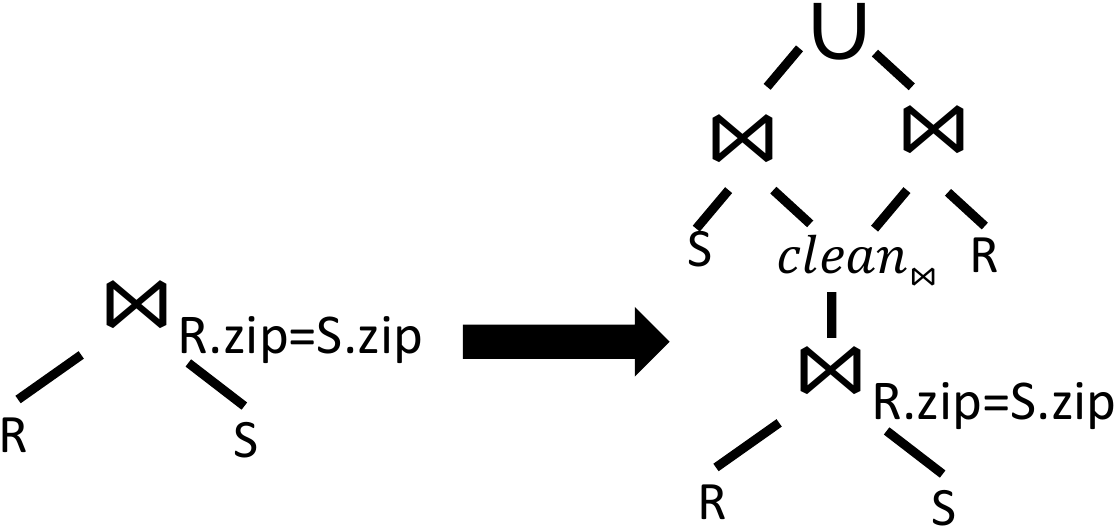}	
	\caption{Before and after injecting $clean_{\bowtie}$ inside a plan with a join over a potentially erroneous attribute. }
	\label{fig:no_filter_plan}
	\vspace{-1.7em}
\end{figure}

%%%%%%%%%%%%%%%%%%%%%%%%%%%ADD%%%%%%%%%%%%%%%%%%%%%%%%%%%%%%%
Fig. \ref{fig:no_filter_plan} shows an example query plan with a $clean_{\bowtie}$ operator.
Given two rules involving the zip code of $R$ and $S$, the planner detects the overlap
of the join operator with the rules, and injects the $clean_{\bowtie}$ operator.
To avoid recomputing the join, $clean_{\bowtie}$ sends only
the new tuples of each relation to the join operator. Thus, the second join corresponds to an incremental
join, which updates the already computed result. The final result consists of the union of the join outputs.

%shows an example query plan which involves cleaning
%operators. Specifically, assume two relations $R$ and $S$, which both involve FD violations. 
%After the join operation, the plan adds a cleaning 
%operator $clean_\Join$ which separately cleans each relation and then updates the join result
%with the additional tuples that result from the cleaning task.

%%%%%%%%%%%%%%%%%%%%%%%%%%%%%%%%%%%%%ADD%%%%%%%%%%%%%%%%%%%%%%%%%

\vspace{-0.5em}
\subsection{Cost-based optimization}
\label{sec:cost_model}

%Performing data cleaning on-demand is beneficial when the queries and the dataset 
%The cost of performing data cleaning on demand is equivalent to the overhead that the cleaning task adds to each query.
In the following, we analyze the offline and incremental cleaning cost, and 
propose a cost model that decides on the optimal placement of cleaning operators in the query plan.

%\subsection{Cost Analysis}
%
%In the following, we analyze the data cleaning cost by taking into consideration different types of rules, as well as
%different types of queries.
%
\vspace{-0.5em}
\subsubsection{Traditional cleaning cost}
The cost of cleaning DCs is divided into the cost of a) error detection, b) data repairing, and, 
	optionally, c) updating the dataset with the correct values.
	For FDs, error detection groups data
	based on the \textit{lhs} of the rule. The complexity of grouping 
	%%%%%%%%%%%%%%%ADD%%%%%%%%%%%%%
	assuming a hash-based algorithm 
	over a dataset of size \textit{n} is \textit{O(n)} \cite{inmenory_dbs}.
	Similarly, for DCs, the cost is \textit{O($n^2$)} since a cartesian product is required,
	%%%%%%%%%%%%%%ADD%%%%%%%%%%
	but is reduced to {\footnotesize $(1+2+..+n)$} since the upper diagonal matrix is checked to avoid re-checking symmetric pairs.
	Data repairing performs multiple scans to compute the candidate values for each error; 
	given $\epsilon$ errors, the cost is \textit{O($\epsilon$ n)}.
	Finally, the update cost is equivalent to an outer join between the dataset and the fixed values; 
	%the cost is $O(n+\epsilon\,p)$
	%where $p$ is the average number of candidate values that each erroneous cell can take.
	the cost is \textit{O(n+$\epsilon$)}.
	%given a dataset with $n$ entities, out of which $\epsilon$ are erroneous,
	The overall cost is: \textit{O(n)/O($n^2$) + O($\epsilon$ n) + O(n+$\epsilon$)}, and can be repeated multiple times
	if many iterations are needed \cite{holistic_dc}.

\vspace{-0.5em}
\subsubsection{Incremental cleaning cost}
\label{subsec:incremental_cost}
We present the incremental cleaning cost by taking into consideration the type of query.
%depends on the query type because
%queries which involve multiple datasets might require re-evaluating join operators in order to ensure that the result of the query
%includes all the candidate tuples. %In the following, we analyze the cost of a single SP query and present the number
%of iterations that join queries might require. 

\noindent\textbf{SP queries:} The cost is equivalent to the cost of computing the correlated tuples plus the traditional cleaning cost.
%the cost of computing the set of correlated tuples, the violation detection cost, the data repairing cost, 
%and the cost of updating the original dataset with the clean tuples.
%\begin{itemize}
%	\item the cost of fetching the set of enhanced tuples, 
%	\item the cost of violation detection,
%	\item the cost of data repairing, 
%	\item the cost of updating the original dataset with the clean tuples.
%	
%\end{itemize}
%Detecting the set of correlated tuples $E(Q)$ requires a traversal of the data.
%For example, given a query answer $A(Q)$, and a functional dependency $FD:X\rightarrow Y$, where $X$ and $Y$
%are sets of attributes, we search through the dataset for tuples with equivalent values in the attributes $X$ and $Y$. The intuition behind
%the enhanced tuples is that we need to bring tuples which do not belong to the result set $A(Q)$, but which are similar to the tuples that belong to $A(Q)$
%in order to perform the violation detection and data repairing. To extract the similar set of tuples, we need to traverse the set of tuples for 
%which there is no information regarding whether they are clean or dirty.
The cost $e_i$ of computing the set of correlated tuples \textit{E(Q)} is \textit{O($u$)}, where $u$ represents the unknown tuples. Given a dataset with $n$ tuples, $u$ is equal to 
$n$ in the first query, but becomes smaller after each query. Specifically, in the $i^{th}$ query, 
the cost is $n-\sum_{i=1}^{q-1} q_i$, where $q_i$ is the size of the result of query $i$.

Error detection and data repairing are applied over the 
%set $A(Q) \cup E(Q)$, that is, over the 
result set \textit{A(Q)} enhanced with the extra tuples \textit{E(Q)}. 
%Given that the size of the result of query $i$ is $q_i$, the number of the extra tuples is $e_i$, 
Thus, the cost of error detection is \textit{O($q_i$+$e_i$)} for FDs.  
For DCs, the incremental cost in the $i^{th}$ query with result size $q_i$
is $n\,q_i/p$. In the worst-case, the incremental cost is the arithmetic
progression (we omit the division by $p$ for simplicity): 

\vspace{-1.5em}
{
	\scriptsize
	\begin{align*}
	& q_1(n-0)+q_2\,(n-q_1)+ ... + q_n(n-\sum_{j=1}^{j=i-1}q_j)=n(q_1+...+q_n -\sum_{i=2}^{i=n}q_i\sum_{j=1}^{j=i-1}q_j\\
	& = n\, \sum_{i=1}^{i=n} q_i - \sum_{i=2}^{n}q_i\,\sum_{j=1}^{j=i-1} q_j \stackrel{({\sum_{i=1}^{i=n} q_i \leq n })}{\leq}  n^2 - \sum_{i=2}^{n}q_i\,\sum_{j=1}^{j=i-1} q_j  \\
	%&\sum_{i=2}^{n}q_i\,\sum_{j=1}^{j=i-1} q_j \leq \sum_{i=1}^{n}q_i\,\sum_{j=1}^{j=i-1} q_j \leq n\, \sum_{i=1}^{i=n} q_i }
	\end{align*}
}
\vspace{-2.9em}

\noindent The worst-case is when {\footnotesize $\sum_{i=2}^{n}q_i\,\sum_{j=1}^{j=i-1} q_j$} is minimized to $0$, which occurs when one query accessing the whole dataset is executed, thus the cost is equivalent to the offline cost.
%Thus, by checking subsets of the matrix with the same level of parallelization, incremental theta-join requires fewer checks overall compared to the offline theta-join
%even when the workload accesses the whole dataset.

For simplicity, we denote the error detection cost as $d_i$
for incremental and $df_i$ for full cleaning. 
Then, given $\epsilon_i$ violated entities, where
$\epsilon_i \leq (q_i+e_i) << n$, the data repairing cost is \textit{O($\epsilon_i$   ($q_i$+$e_i$))}, since 
it checks for each error the enhanced tuples instead of checking the whole dataset.

The last step involves updating the original dataset with the fixed tuples stemming from cleaning the query result. The update performs
a left-outer-join between the dataset and the clean result. Since the clean result contains
probabilistic values, the update depends on the number of candidate values that the erroneous value might take.
More specifically, assuming a partially probabilistic dataset at query $i$ with $\sum_{j=1}^{j=i-1} \epsilon_{ij}$ probabilistic values, the update cost is: 
{\textit{O($n - \sum_{j=1}^{j=i-1} \epsilon_{ij} + \sum_{j=1}^{j=i-1} \epsilon_{ij}\,p + \epsilon_i\,p$)}}, where $p$ is the size of each value. In total, the incremental cleaning
cost is:

\vspace{-1.2em}
{
	\footnotesize
	\begin{flalign*}
	\label{eq:cost_model}
	n-\sum_{i=1}^{q-1} q_i +d_i + \epsilon_i \cdot (q_i+e_i) + n - \sum_{j=1}^{j=i-1} \epsilon_{ij} + \sum_{j=1}^{j=i-1} \epsilon_{ij}\,p + \epsilon_i\,p
	\end{flalign*}
}
\vspace{-1em}

In the case of multiple rules, the cost differs in the error detection
part, since it requires one iteration per rule. The computation of the probabilities for
the erroneous entities is equivalent to the single rule case because it operates over the detected errors, 
regardless of the number of violated rules.

\noindent\textbf{Join queries: }
%%%%%%%%%%%%%%%%%%%%ADD%%%%%%%%%%%%%
The aforementioned cost represents the \texttt{query and clean} cost of each individual table
that participates in a join. 
However, a join involves the additional cost
of updating the join result. 
Therefore, we need to measure the maximum number of iterations. 
We apply the cost formula for each dataset that participates in the join,
and then we add the incremental join cost which takes place between the extra tuples $e_i$
of one relation with the set of tuples $n$ of the other relation: $(n + e_i)$.
We use the formula separately in each dataset, because each dataset has different characteristics, that is
different number of violations, different level of correlation among the entities, and finally the
query has a different selectivity in each dataset. 
% the cost of incrementally cleaning join queries requires twice the cleaning cost mentioned in (1).

\subsubsection{Incremental cleaning versus Full cleaning}
\label{subsec:cost_model}

The decision between incremental or full cleaning depends on whether the overhead induced by the cleaning task in each query 
is smaller in total than applying the full cleaning followed by the execution of the queries.
To estimate the costs, we employ a cost model that decides on the optimal strategy.

Cleaning at query time without considering the relaxation and the update cost, is more efficient overall than executing them over the whole dataset.
Consider an unknown query workload consisting of $q$ queries. 
In the case of FDs, the cost $\sum_{i=1}^{q}\epsilon_i(q_i+e_i) \leq \epsilon\,n$ since, $q_i$
and $e_i$ are complementary, thus, their total number of tuples is smaller than the total dataset.
In addition, for DCs, the incremental error detection cost is smaller than the cartesian product.
However, the cost of enhancing the query result and updating the dataset after each query might exceed the full cleaning
cost. Thus, we decide on cleaning the query result or the remaining dirty part of the dataset based on the following inequality. 
In the offline cost we also add the query execution cost, which is $q\,n$:

\vspace{-1.1em}
{	\footnotesize
	\begin{flalign*}
	\sum_{i=1}^{i=q}(n-\sum_{j=1}^{j=i-1}q_j+d_i+\epsilon_i\,(q_i+e_i)+n-\sum_{j=1}^{j=i-1}\epsilon_{ij}+\sum_{j=1}^{j=i-1}\epsilon_{ij}\,p+\epsilon_i\,p) \\
	\leq qn + df_i + \epsilon\,n + n+\epsilon\,p	
	\end{flalign*}
}
\vspace{-1.1em}

%\noindent In the worst case, the query workload accesses the whole dataset, thus, $\sum_{i=1}^{i=q}q_i=n$, $\sum_{i=1}^{i=q}\epsilon_i=\epsilon$,
%and $\sum_{i=1}^{i=q}e_i =T \leq n$. 
The inequality can be simplified to the following one:
%\vspace{-0.7em}
{	\scriptsize
	\begin{align*}
	& q\,n - \sum_{i=1}^{i=q}\sum_{j=1}^{j=i-1}q_j + d_i + \sum_{i=1}^{i=q}\epsilon_i\,q_i + \sum_{i=1}^{i=q} \epsilon_i\,e_i - \sum_{i=1}^{i=q}\sum_{j=1}^{j=i-1}e_{ij} +
	p\sum_{i=1}^{i=q}\sum_{j=1}^{j=i-1}e_{ij} \\
	& \leq df_i + \epsilon\,n + n %\Longleftrightarrow  \\ 
	%& q\,n - \sum_{i=1}^{i=q} \sum_{j=1}^{j=q-1}q_j + n + n + \epsilon\,s + \epsilon\,e + q\,n\,p + \epsilon\,p
	\end{align*}
}
\vspace{-1em}

%%%%%%%%%%%%%%%%%%%%%%%%%%%%%%%%%%%%%%ADD%%%%%%%%%%%%%%%%%%%%%%%%%%%%%%%%
For example, when $q=1$, and $\sum_{i=1}^{n}q_i=n$, then $q_1=n$, $e_1=0$ since the query accesses the whole dataset, therefore there are no extra tuples. Thus, the cost
corresponds to the full cleaning case and the inequality becomes:

\vspace{-1.2em}
{\footnotesize
	\begin{align*}
	& n + n + \epsilon\,n + 0 - 0 + 0 \leq \epsilon\,n + 2n \Longleftrightarrow \epsilon\,n \leq \epsilon\,n \\
	\end{align*}
}
\vspace{-2.5em}

%Since, the formula (1) is always positive, it is maximized when $q=n$. In that case $\forall i, q_i=1$:
%Thus, in the end we would have:
%\vspace{-0.3em}
%{\tiny
%\begin{align*}
%\epsilon \geq  n^2-n/2\,n-2-T+n^2-n-n^2p+n\,p
%\end{align*}
%}
%\vspace{-0.3em}

We observe from the inequality that in the case of general DCs, since \textit{p} increases when the selectivity is high, then
Algorithm \ref{algo:dc} will also decide to examine the whole cartesian product due to predicting low accuracy.
For FDs, we decide on the cleaning strategy while executing the queries based on the inequality.
We estimate the number of erroneous values $\epsilon$, as well as the number of candidate values $p$ using statistics.
To approximate $\epsilon$ and $p$, we precompute the group by based on the $lhs$ and the $rhs$ of the FD rules respectively.
%and b) a histogram to estimate the selectivity of the theta-join, following the approach of \cite{mr_joins,inequality_joins}.

%Therefore, using the statistics that pre-compute the number of violations, one can estimate whether the components 
%that do not involve the characteristics of the query workload are greater than the cost of full cleaning.
%Specifically, in the case where the number of candidate values $p$ is close to $n$, then \system decides to fully clean
%the dataset.
%A similar case holds in the case of join queries with the difference that we include the number of iterations in the inequality.

%\noindent \textbf{Summary}: In general, full cleaning is preferred when the erroneous entities get multiple candidate values, 
%that is the value $p$ becomes comparable to the size of
%dataset. 
%This happens when the probabilities $P(lhs|rhs)$ or $P(rhs|lhs)$ become very small. 

%To compute the probability of having an extra iteration, we need to compute the probability of having an extra tuple in the join due to the cleaning
%task.

%%\input{cost_model}
%%\input{accuracy}
\section{A system for query-driven \\ data cleaning}
\label{sec:system}
%\vspace{-0.5em}

\begin{figure}[t]
	\centering
	\includegraphics[width=.25\textwidth]{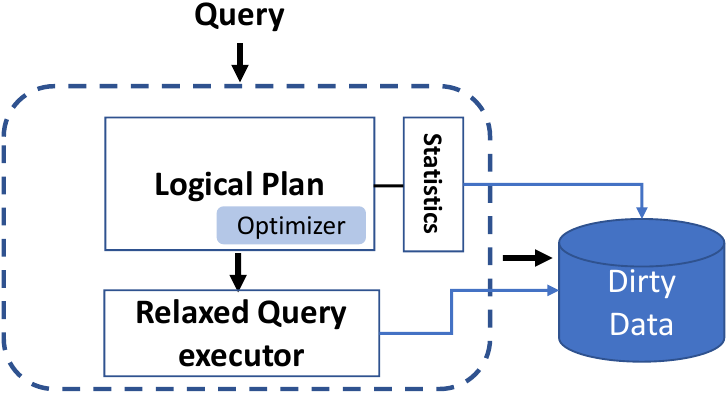}		
	\caption{The architecture of \systemNoindent.}
	\label{fig:system}
	\vspace{-1.7em}
\end{figure}

%We validate our approach by building \systemNoindent, a query-driven cleaning
%engine over Spark.
Fig. \ref{fig:system} shows the architecture of \systemNoindent, that is a query-driven cleaning
engine over Spark.
Given a query and a dirty dataset, \system uses two processing levels to provide correct results.

In the first level, \system maps the query to a logical plan 
that comprises both query and cleaning operators. To optimally
place each operator, the logical plan
takes into consideration
the type of query and the constraints. We implement the cleaning-aware logical plan by injecting cleaning operators
before/after the corresponding filter and join operators at the RDD level of Spark. \system extracts the attributes
of the query operators and checks if they overlap with the provided constraints.
To apply the cost-based optimizations, \system collects statistics by pre-computing the size
of the erroneous groups. Then, when checking the condition of each query, it evaluates
whether the inequality of Section \ref{subsec:cost_model} holds. Hence, at the logical level \system decides
whether to place the cleaning operator before or after the query operator.

Finally, \system executes the logical plan by cleaning the result of each query operator that is affected by the rules. 
We implement $clean_{\sigma}$ and $clean_{\bowtie}$ as extra operators inside Spark RDD.
The operators take as input the query result, relax it, and detect for violations.
Then, given the detected violations, \system transforms the query result into a probabilistic
result by replacing each erroneous value with the set of values that represent candidate fixes.
\system also accompanies each candidate value with the corresponding probability of being a fix of the erroneous cell. 
After cleaning each query result, the system isolates the changes made to the erroneous
tuples and accordingly updates the original dataset. By applying the changes after each
query, \system gradually cleans the dataset.

\section{Experimental Evaluation}
\label{sec:eval}
\vspace{-0.5em}

The experiments examine the benefits stemming from the optimizations that \system allows,
and show how \system performs compared to the state-of-the-art offline cleaning approach.

\noindent\textbf{Experimental Setup.}
All experiments run on a 7-node cluster equipped with 2$\times$Intel(R) Xeon(R) Gold 5118 CPU (12 cores per socket @ 2.30GHz), 64KB of L1 
cache and 1024KB of L2 cache per core, 16MB of L3 cache shared, and 376GB of RAM. 
On top of the cluster runs Spark 2.2.0 with 7 workers, 14 executors, each using 4 cores and 150GB of memory.

We compare a single-node execution of \system with Holoclean \cite{holoclean} 
as it is, to our knowledge, the only currently available probabilistic system
for repairing integrity-constraint violations.
In the absence of a scale-out probabilistic cleaning system, we 	
compare \system with our own offline implementation over Spark;
it combines
the optimizations of the state-of-the-art error detection and probabilistic repairing systems.
Our offline implementation is an optimized implementation that detects FD and DC errors, and it provides probabilistic repairs.
Error detection follows the optimizations of BigDansing~\cite{bigdansing} for FDs; it applies a group-by, instead of an expensive self-join. DC error detection efficiently partitions the cartesian product  using the optimized theta-join approach \cite{mr_joins}.
Directly comparing \system with BigDansing would be unfair to BigDansing because BigDansing applies inference in order to compute the correct value, whereas
\system computes probabilistic repairs.
For data repair, to restrict the 
domain of candidate values for each erroneous cell, we employ an alternative to Holoclean's pruning optimization \cite{holoclean};
%For the partitioning optimization, we partition tuples based on the 
%the $lhs$ of each violating FD group. 
we exploit
the co-occurrences of the attribute values of the erroneous tuple with the attribute values of other tuples. Hence, similarly to \systemNoindent, the domain
of the erroneous $rhs$ of tuple $t$ correspond to the $rhs$ of the tuples that share the same $lhs$ with $t$. Similarly for the erroneous $lhs$.
%Directly comparing against Holoclean and BigDansing is unfair because a) Holoclean is
%a single-node system whereas \system is scale-out, and b) both Holoclean and BigDansing apply an inference task
%to compute the correct value, whereas \system only lists the set of candidate values.

%%%%%%%%%%%%%%%%%%%%%%%%%%%%%%%%ADD - REPLACE%%%%%%%%%%%%%%%%%%%%%%%%%%%%%%%%%%%%%%%%%%%%%%%%
The workload involves SP, SPJ, and group-by queries in the presence of one or more DCs.
We evaluate the workload over a synthetic benchmark and three datasets derived from real-world data entries.
Specifically, we use the Star Schema Benchmark (SSB) \cite{ssb}, the hospital
dataset \cite{holoclean}, the Nestle dataset and
a dataset with air quality data~\cite{epa-historical-air-quality}. 

We choose the SSB dataset to test the applicability of \system over a benchmark designed for data warehousing applications.
%We execute the workload over the \textit{lineorder} and \textit{supplier} tables. 
We use multiple versions of the \textit{lineorder} table by varying the cardinality of the \textit{orderkey} and \textit{suppkey} attributes;
we construct different versions by varying the number of distinct \textit{orderkeys} from 5K to 100K, and the number of distinct \textit{suppkeys}
from 100 to 10K. 
To measure the worst-case scenario, we add errors to all orderkeys by randomly editing 10\% of the suppliers that correspond to each \textit{orderkey}.
Our error generation is similar to BART~\cite{bart} with the difference that we also add errors using uniform distribution to evenly distribute the errors across the dataset, thereby affecting all queries. 
The errors that we inject are detectable by the constraints that we evaluate.
The size of \textit{lineorder} table is 60MB in the original version, and ranges from 110MB to 2.6GB in the probabilistic version. %, after accessing the whole dataset.
To evaluate cases with fewer violations, we construct datasets with 20\%, 40\%, 60\% and 80\% of erroneous
orderkeys. %The resulting datasets are 20\%, 40\%, 60\% 80\% and 100\% erroneous.
The size of the probabilistic version of those datasets is 250MB, 560MB, 1.3GB, and 1.8GB.

The hospital dataset \cite{holoclean} comprises information about US hospitals. It contains 19 attributes, and is 5\% erroneous.
We use two versions with 1K and 100K entries, and sizes 300KB, 25MB respectively. The probabilistic versions have size 360KB and 26MB respectively.
We use hospital
to evaluate accuracy since its clean version exists. 
%We also use it to compare against Holoclean since it was the dataset for which Holoclean managed to terminate by successfully fixing the dataset.

%%%%%%%%%%%%%%%%%%%%%%%%%%ADD%%%%%%%%%%%%%%%%%%%%%%%%%%
The Nestle dataset includes information about food and drink products. Each product contains 19
attributes and involves dirty categories for product materials.
We scale up the dataset by randomly adding duplicate entities from the domain of each attribute.
We also add extra errors by randomly editing 10\% of the \textit{category} attribute values that correspond to each \textit{material}.
We use a 20MB and a 200MB version which contain 95\% of conflicting entities. The size of the datasets in the probabilistic
version is 40MB and 500MB respectively.

The historical air quality dataset \cite{air_quality_analysis} contains air quality measurements for the U.S. counties. We use a subset
of the hourly measurements in which we add errors to the FD \textit{$\phi$:county\_code,state\_code$\rightarrow$county\_name}.
We edit 10\% of the \textit{county\_names} that correspond to a \textit{county\_code,state\_code}.
We add the errors to the non-frequent \textit{county\_code,state\_code} pairs. 
We use two versions with 0.001\% and 0.003\% errors respectively, which produce 30\% and 97\% violations respectively, the size of which is 2GB in the original version and 3.1GB and 4GB in
the probabilistic versions.

We measure response time and accuracy (when applicable). 
Response time is the time to respond to the query, perform the cleaning task by providing probabilistic fixes, and update the dataset.
For accuracy, we measure \textit{precision} (correct updates/total updates) and \textit{recall} (correct updates/total errors). 

\vspace{-0.3em}
\subsection{SP queries response time}
%\vspace{-0.3em}
\label{subsec:sp_eval}

This section shows how \system performs compared to offline cleaning given a workload of SP queries.
We measure the cost of both approaches given a) a FD, b) two overlapping
FDs, and c) a DC. 
We evaluate all cases over SSB, by executing queries requesting information for a specific supplier/order, or for suppliers/orders in a given range.
In all FD experiments, \system outputs the same results with the offline approach.
%We also evaluate the aforementioned approaches over data with 
%increasing number of violations in order to capture all possible cases of dirty data. We execute a set of 50 SP 
%queries over the SSB dataset, which are of the following form:
%\vspace{-0.4em}
%\begin{minted}[fontsize=\footnotesize]{sql}
%SELECT orderkey, suppkey
%FROM lineorder
%WHERE [suppkey=X | suppkey > X AND suppkey < Y] 
%\end{minted}
%%\vspace{-0.4em}
%\noindent where \texttt{X, Y} are constant values. Queries request information for a specific supplier, or for suppliers in a given range. 

%%%%%%%%%%%%%%%%%%%%%%%%%%%%ADD-REPLACE%%%%%%%%%%%%%%%%%%%%%%%%%%
\noindent\textbf{Single FD with varying selectivity of rule attributes.}
We examine how the \textit{orderkey} and \textit{suppkey} selectivity affects the response time of the cleaning task.
%We use multiple versions of the lineorder dataset by varying the number of distinct values for the \textit{orderkey} and \textit{suppkey}. Specifically,
We use three versions of lineorder with 5K, 10K, and 100K distinct \textit{orderkey} values respectively, and three versions
with 100, 1K, and 10K distinct \textit{suppkey} values respectively. 
We use these selectivities since they involve extreme response time cases depending on the query.
We clean violations of rule \textit{$\phi$:orderkey$\rightarrow$suppkey}.
We consider the worst-case scenario where each \textit{orderkey} participates in a violation. We execute 50 non-overlapping queries, each with
selectivity 2\%. The workload accesses the whole dataset. 

%%%%%%%%%%%%%%%%%%%ADD%%%%%%%%%%%%%%%%%%%%%
\begin{figure}[t]
	\centering
	%%%%%%%%%%%%%%%%%%%%%%%%%%%ADD%%%%%%%%%%%%%%%%%%%%%%%%%%
	\begin{minipage}{0.21\textwidth}
		\includegraphics[width=\textwidth]{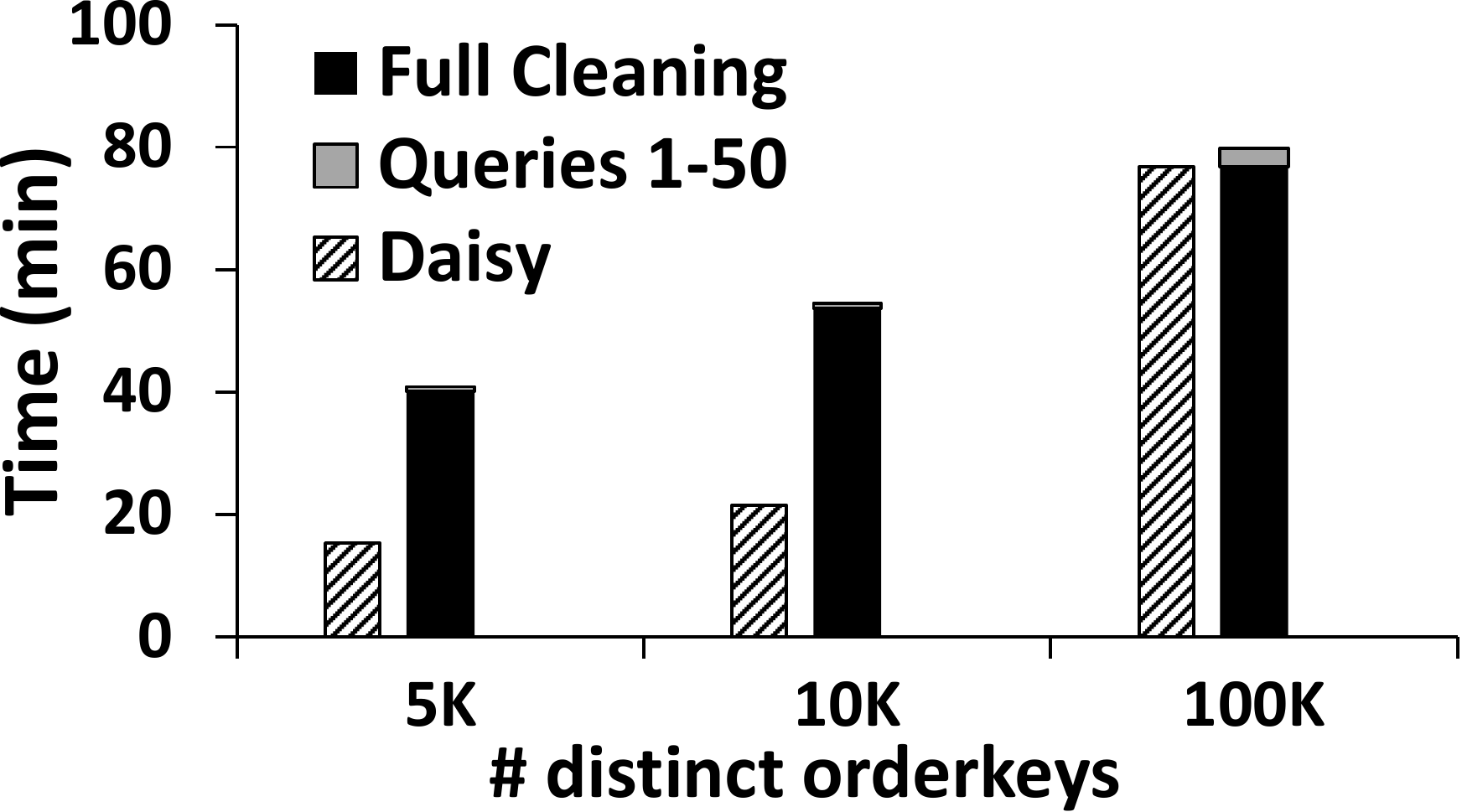}
		%	\vspace{-0.5em}	
		\caption{Cost when varying orderkey selectivity.}
		\label{fig:sp_all}
	\end{minipage}
	\hspace{1em}
	\begin{minipage}{0.21\textwidth}
			\includegraphics[width=\textwidth]{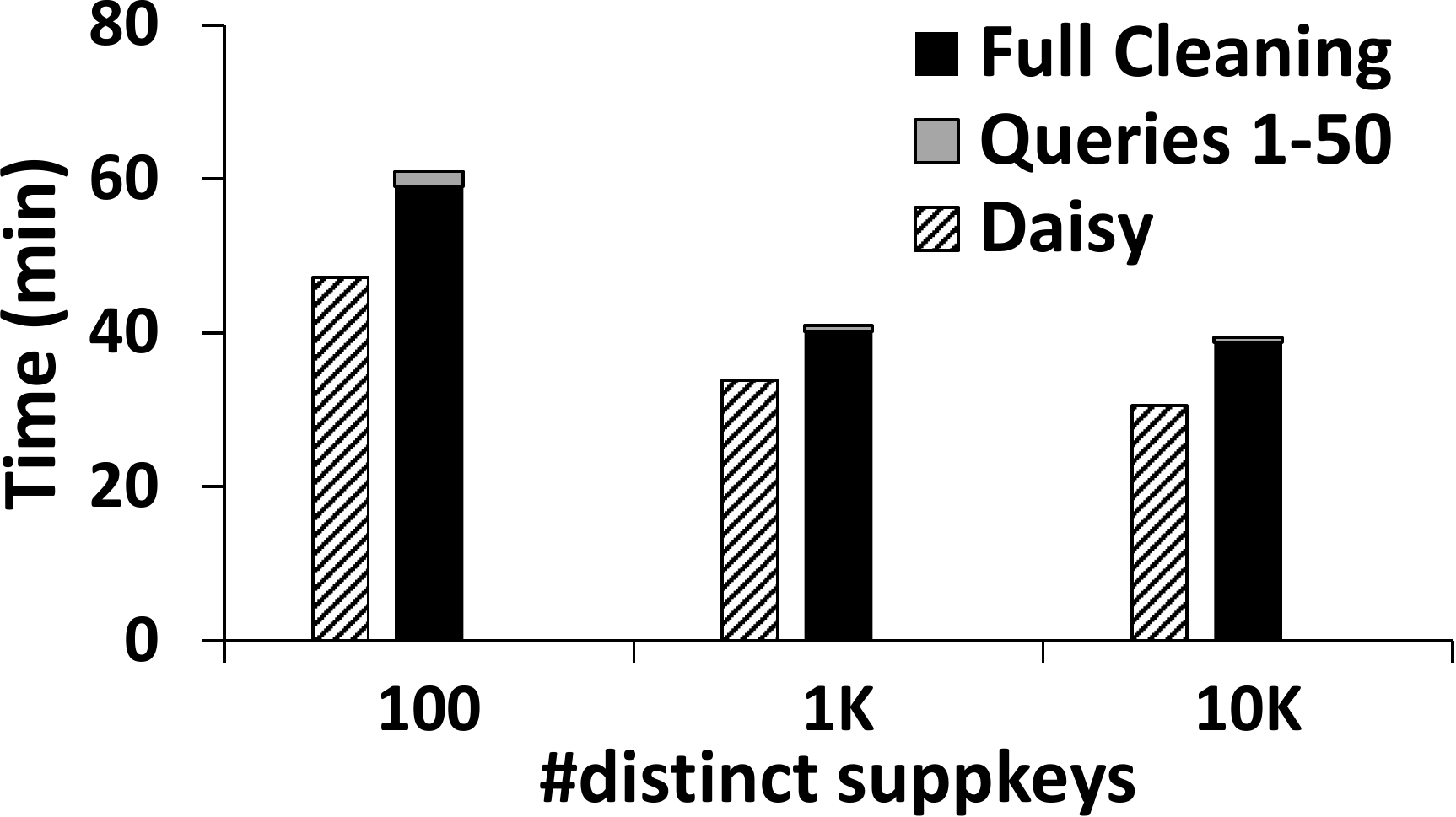}
			\vspace{-1.2em}	
			%		\captionsetup{width=1.4\textwidth}
			\caption{SP Cost when varying suppkey selectivity.}
			\label{fig:sp_lhs_all}
	\end{minipage}	
	\vspace{-1.8em}	
\end{figure}

Fig. \ref{fig:sp_all} shows the response time of \system and full cleaning when varying the \textit{orderkey} selectivity. 
To maintain a fixed query selectivity, queries contain range filters over the $rhs$ of $\phi$.
We observe that as the selectivity increases, the response time of both approaches increases.
However, on average, \system is $\sim2\times$ faster than the offline approach.
The difference
is due to the fact that when combining cleaning with querying, query result relaxation 
restricts the number of comparisons required to repair the erroneous tuples by computing the correlated tuples.
On the other hand, the
offline approach traverses the dataset for each erroneous value, 
to compute the candidate values. We also observe that as the selectivity increases, the difference between
the two approaches decreases because each erroneous cell ends up having more candidate values 
thereby increasing the value $p$ of the inequality of Section~\ref{subsec:cost_model}.
%the update cost for \system after each query.

Fig. \ref{fig:sp_lhs_all} shows the response time of \system and offline cleaning when varying \textit{suppkey} selectivity. 
To maintain a fixed query selectivity, queries contain range filters over the $lhs$ of $\phi$.
\system is faster
despite the transitive closure it requires to detect the correlated values. 
The difference
is due to the fact that when combining cleaning with querying, query result relaxation 
restricts the number of comparisons required to repair the errors by computing the correlated tuples.
On the other hand, the
offline approach traverses the dataset for each erroneous value 
to compute the candidates.
When \textit{suppkey} selectivity is smaller, the cost becomes higher since each erroneous \textit{suppkey} might
match with multiple \textit{orderkeys}, thereby increasing the number of candidate values.

\begin{figure}
	\begin{minipage}{0.21\textwidth}
		\includegraphics[width=\textwidth]{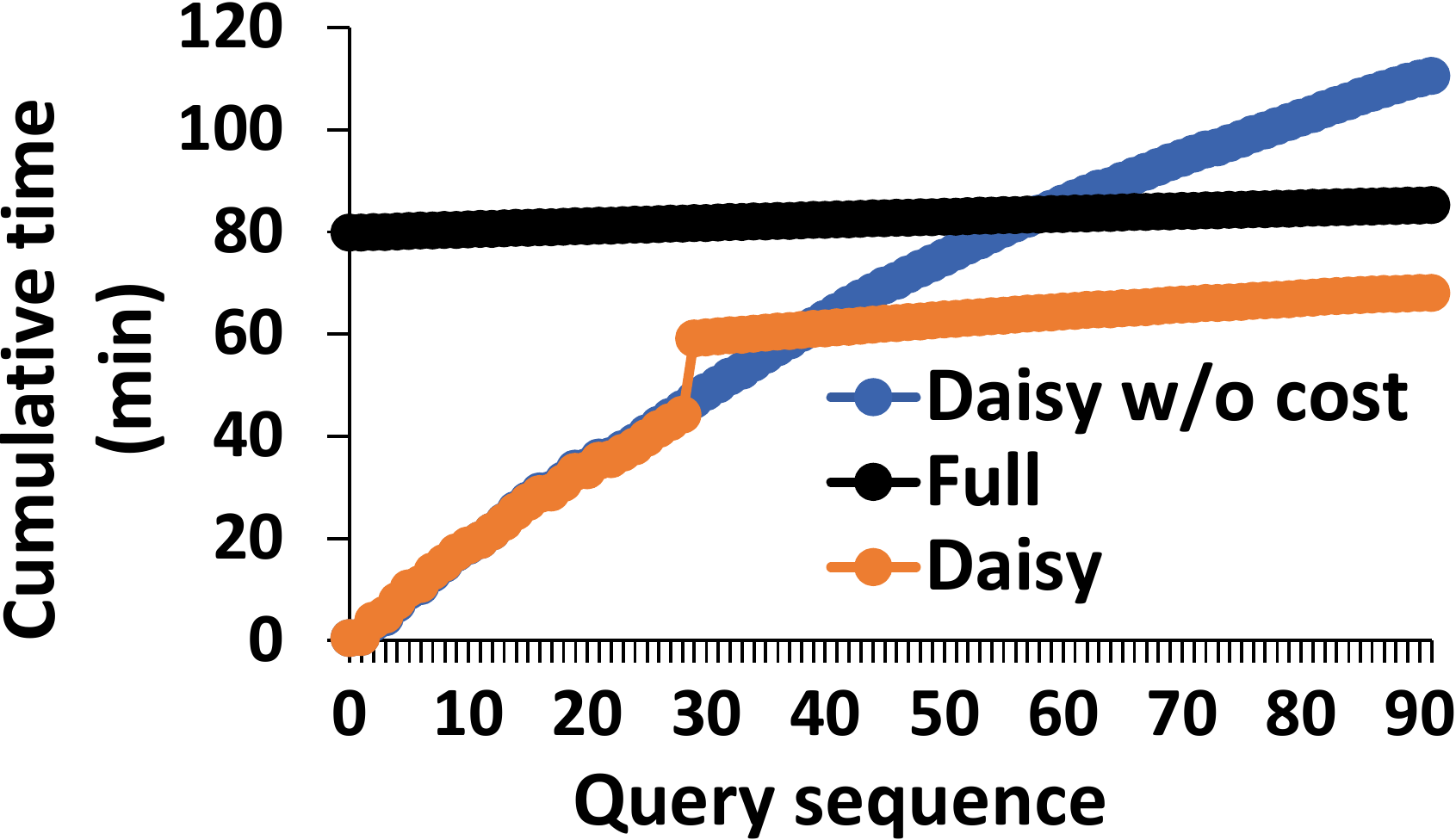}
		\vspace{-1.2em}	
		%		\captionsetup{width=1.4\textwidth}
		\caption{Switching from incremental to full cleaning.}
		\label{fig:sp_random}
	\end{minipage}
	\hspace{1em}	
	\begin{minipage}{0.21\textwidth}
		\includegraphics[width=\textwidth]{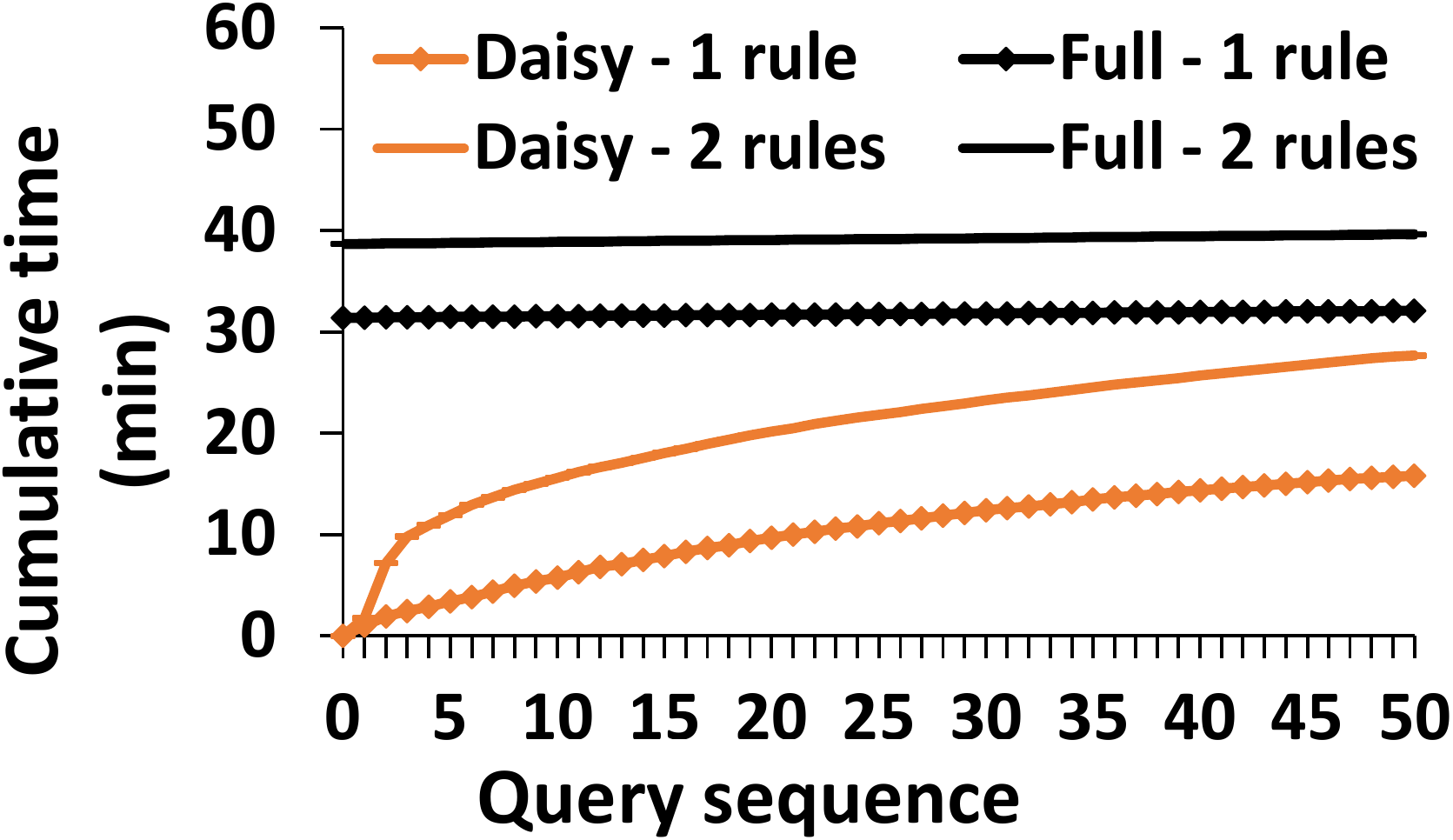}		
		\vspace{-1.2em}	
		\caption{Cost when increasing number of rules.}
		\label{fig:sp_multirule}
		%		\vspace{-0.2em}	
	\end{minipage}	
	\vspace{-2em}	
\end{figure}

Fig. \ref{fig:sp_random} evaluates the scenario in which applying cleaning offline outperforms incremental cleaning (\system without the cost model).
We execute 90 queries over the lineorder version with 100K distinct \textit{orderkeys}. The queries are non-overlapping, they involve
equality and range conditions, and have random selectivities.
Cleaning the whole dataset is more efficient in this case because the suppkey selectivity is low
compared to the orderkey,
thus each suppkey appears with multiple orderkeys throughout the dataset. Thus, a violating suppkey takes multiple
candidate values, thereby increasing the update cost shown in the inequality of Section \ref{subsec:cost_model}.  
%Thus, the overhead of the cleaning task in each query is high.
Still, we observe that overall, \system outperforms both the incremental as well as the offline cleaning.
\system initially applies data cleaning incrementally, and then, by evaluating the total cost after each
query, switches strategy and applies the cleaning task over the rest of the dataset. The total cost is
lower than the offline approach because cleaning is applied over the remaining dirty part of the dataset.
%
%We observe
%that in the 10K case where the selectivity is smaller, \system follows the incremental cleaning approach which outperforms the full
%cleaning approach even though both approaches need to access the whole dataset.
%and how \system decides on the cleaning strategy based on the cost model.

\noindent\textbf{Single rule vs. Multiple rules.}
In this experiment, we measure the response time in the presence of rules with overlapping attributes. We construct
the dataset by joining lineorder with suppliers. The end result is a 67MB dataset in its raw form,
and 2.8GB in its probabilistic form. We evaluate rules \textit{$\phi$:orderkey$\rightarrow$suppkey}
and \textit{$\psi$:address$\rightarrow$suppkey}; the \textit{address} appears after joining the tables.
The workload consists of 50 non-overlapping queries which access the whole dataset.

Fig. \ref{fig:sp_multirule} shows the response time in the case where
we examine only rule $\phi$ compared to examining both $\phi$ and $\psi$.
We observe that in both \system and the offline approach, response time increases when we clean errors of both rules instead of one, due to the extra work required for $\psi$. When \system executes the queries, it identifies 
the corresponding correlated tuples for both rules. Then, \system fixes the errors based on the correlated tuples.
Initially, the difference between one and two rules is $\sim3.5x$ but then
drops to $\sim1.5x$ as we clean more data.
On the other hand, offline
cleaning separately fixes the errors of the \textit{address} and \textit{orderkey} since there
might be different tuples involved in the violation of $\phi$ than those involved in $\psi$. Thus, offline cleaning needs more traversals over the data.

%\begin{figure}[t]
%	\centering
%	\includegraphics[width=.45\textwidth]{figures/multirule}		
%	\caption{Response time when increasing the number of functional dependencies.}
%	\label{fig:multirule}
%	\vspace{-0.2em}	
%\end{figure}

%\begin{figure*}[t]
%	\begin{minipage}{0.3\textwidth}
%		\centering
%		\includegraphics[width=\textwidth]{figures/join_5}		
%%		\vspace{-1em}		
%		\caption{Response time for join queries.}
%		\label{fig:join}	
%%		\vspace{-1em}		
%	\end{minipage}
%	\hspace{6em}
%	\begin{minipage}{0.3\textwidth}
%		\centering
%		\includegraphics[width=\textwidth]{figures/join_random}		
%		%		\vspace{-0.5em}
%%		\vspace{-1em}		
%		\caption{Response time for workload with SP and join queries.}
%		%	\vspace{-0.5em}
%		\label{fig:join_random}	
%%		\vspace{-1em}				
%	\end{minipage}		
%\end{figure*}

%%%%%%%%%%%%%%%%%%%%%%%%%%%%%ADD%%%%%%%%%%%%%%%%%%%%%%%%%%%%%%%%%%%%%%%%%%%%5
\noindent\textbf{Increasing number of violations.}
In this experiment we evaluate \system as we vary the number of violations. 
%by increasing the number of orderkeys that participate in a violation. 
Specifically, we vary the erroneous orderkeys from 20\% to 80\%. We use the same
query workload consisting of 50 non-overlapping SP queries with selectivity 2\%.

\begin{figure}[t]
	%%%%%%%%%%%%%%%%%%%%%%%%%%%ADD%%%%%%%%%%%%%%%%%%%%%%%%
	\begin{minipage}{0.21\textwidth}
		\centering
		\includegraphics[width=\textwidth]{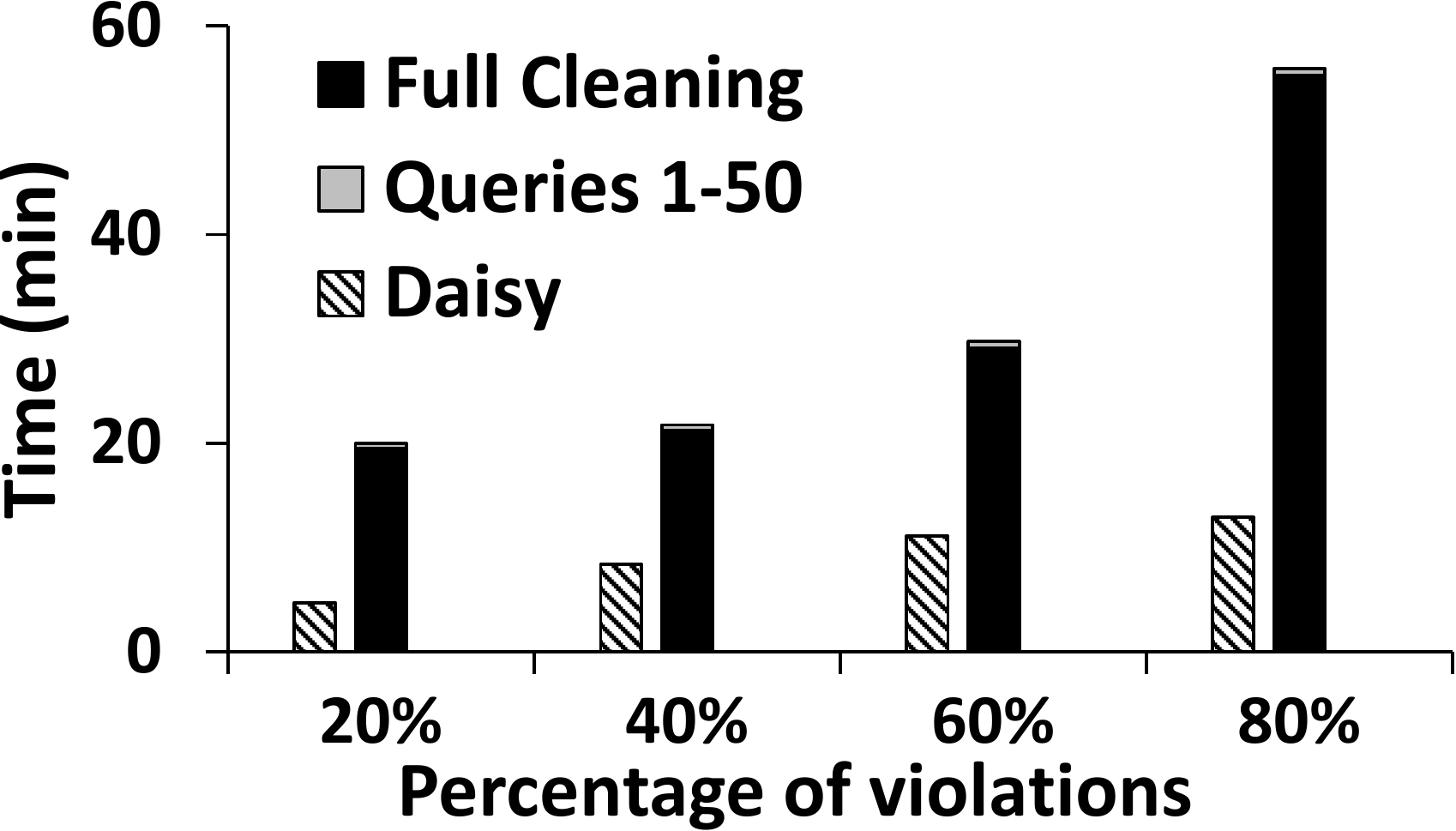}		
		%		\vspace{-0.5em}
		\caption{Cost with increasing number of violations.}
		%	\vspace{-0.5em}
		\label{fig:sp_increasing_vio}
		\vspace{-1.7em}	
	\end{minipage}		
	\hspace{2em}
	\begin{minipage}{0.21\textwidth}
		\vspace{-0.3em}
		\centering
		\includegraphics[width=\textwidth]{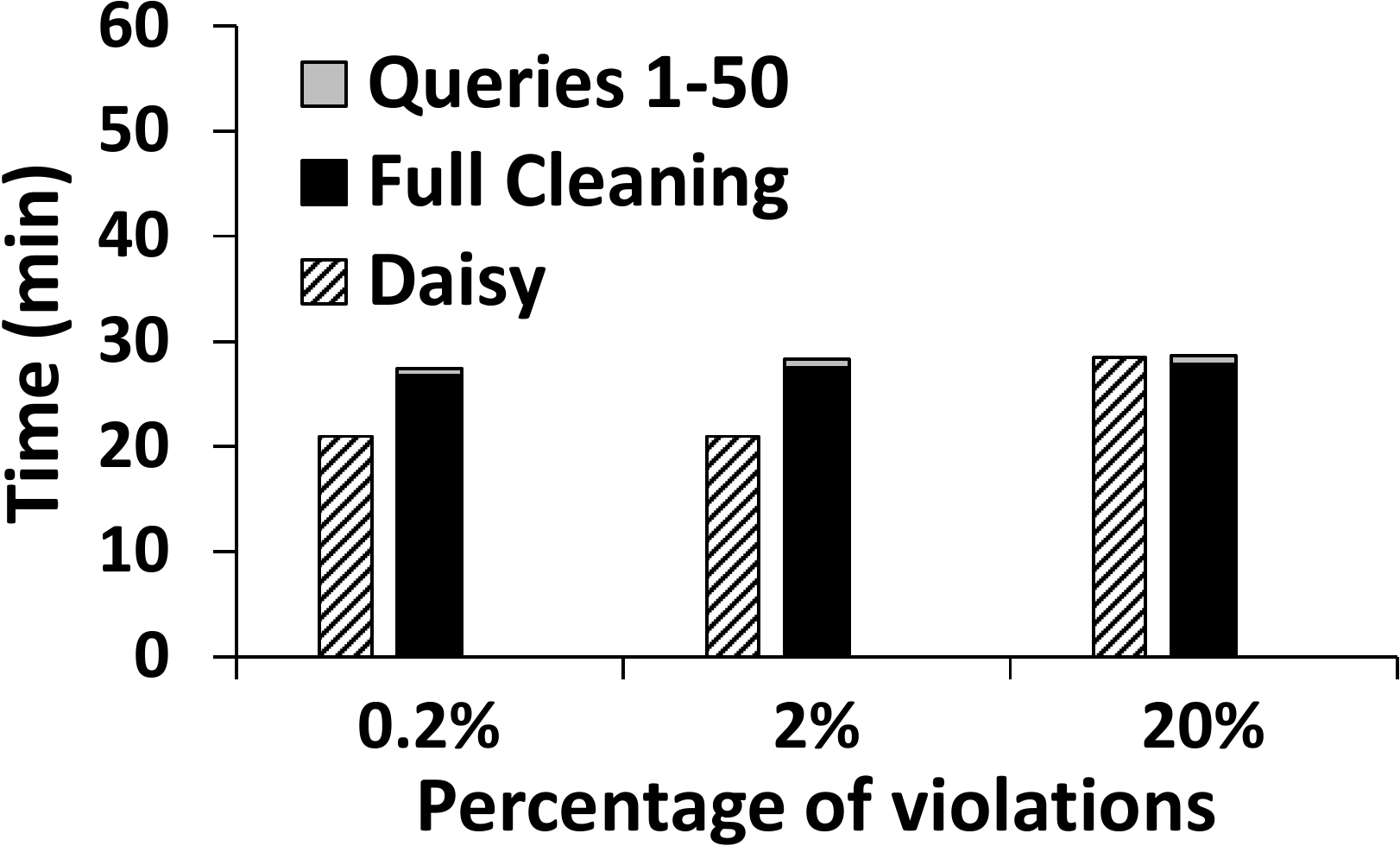}		
		%		\vspace{-0.5em}
		\caption{Cost for DCs with inequality conditions.}
		%	\vspace{-0.5em}
		\label{fig:sp_ineq}	
		\vspace{-2em}
	\end{minipage}	
\end{figure}

Fig. \ref{fig:sp_increasing_vio} shows that in all cases \system outperforms the offline approach
regardless of the number of erroneous entities. \system is faster due to the statistics
that it precomputes to prune unnecessary checks. The statistics comprise 
the orderkeys that participate in a violation; it precomputes a group by based on the orderkey
and calculates the size of each group. Then, at query time, when it accesses a specific orderkey, it checks whether it belongs 
to a dirty group. Thus, \system avoids detecting violations when the entity does not belong to the list of dirty values.
We also observe that as errors increase, the difference between the two approaches
is more significant. The difference stems from the fact that in the case of
full cleaning, the number of iterations over the dataset is proportional to the number of detected
erroneous groups in order to compute the probabilities of each candidate value. On the other hand, depending on the values that the query accesses, \system traverses the data
once and brings the correlated tuples that correspond to multiple erroneous groups at the same time.
Thus, as we increase the number of erroneous groups, offline cleaning performs
more traversals, and thus becomes slower.

%\begin{figure}
%	\begin{minipage}{0.21\textwidth}
%		\includegraphics[width=\textwidth]{figures/sp_multirule}		
%		\vspace{-1.2em}	
%		\captionsetup{width=1.1\linewidth}
%		\caption{Cost when increasing number of rules.}
%		\label{fig:sp_multirule}
%		%		\vspace{-0.2em}	
%	\end{minipage}	
%	\hspace{2em}
%	\begin{minipage}{0.21\textwidth}
%		\includegraphics[width=\textwidth]{figures/sp_ineq}		
%		\vspace{-1.2em}	
%		\captionsetup{width=1.1\linewidth}
%		\caption{\red{Cost for DCs with inequality conditions.}}
%		%	\vspace{-0.5em}
%		\label{fig:sp_ineq}	
%	\end{minipage}	
%	\vspace{-2.5em}	
%\end{figure}

%%%%%%%%%%%%%%%%%%%%%%%%%%%%%%ADD%%%%%%%%%%%%%%%%%%%%%%%%%%%%%%%%%

\noindent\textbf{Denial constraints.}
In this experiment, we evaluate the cost, given rules with inequality predicates.
We consider rule {\small \textit{$\forall$$t_1$,t$_2$$\lnot$ ($t_1$.extended\_price$<$$t_2$.extended\_price\&$t_1$.discount$>$$t_2$.discount)}}.
We check the rule over the lineorder table in which we inject errors by editing the \textit{discount} value of 10\% of entries. 
We simulate real-world scenarios that, unlike high selectivity inequality joins, 
contain a few dirty values that cause inconsistencies.
We construct
three versions with 0.2\%, 2\%, and 20\% violations,
by modifying the errors that the dirty values induce.
We execute 60 SP non-overlapping range queries that access the whole dataset.

Fig. \ref{fig:sp_ineq} shows the response time of both \system and the optimized offline approach. In the 0.2\% and 2\% versions, 
\system is $1.3\times$ faster, as it prunes both the partitions and the subset of the partitions that must be checked.
The result is 99\% and 80\% accurate, respectively, compared to the offline case.
In the 20\% case, \system predicts 23\% accuracy by using the statistics, hence it decides to clean the whole
dataset and is 100\% accurate and has the same response compared to the offline case.
Specifically, in the 20\% case, the dirty values are spread across different partitions and contain outlier values that affect the result.
Therefore, this case justifies the need for checking the whole matrix to provide an accurate result.

\subsection{SPJ queries response time}
\label{subsec:join_eval}

This section demonstrates how \system performs when
Join queries appear in the workload. 
We execute 50 join queries over lineorder and suppliers.
The lineorder table violates rule \textit{$\phi$:orderkey$\rightarrow$suppkey}, and
the suppliers violates rule \textit{$\psi$:address$\rightarrow$suppkey}.
The queries contain a filter on lineorder, and then join it with suppliers.
The workload accesses
the whole lineorder dataset. %The syntax of each query is the following:
%\vspace{-0.4em}
%\begin{minted}[escapeinside=||,fontsize=\footnotesize]{sql}
%SELECT lo.orderkey, s.suppkey
%FROM lineorder lo, suppliers s
%WHERE lo.suppkey=[X] AND lo.suppkey = s.suppkey
%\end{minted}
%\vspace{-0.4em}

%\noindent where the value \texttt{X} is a constant that requests information for a specific supplier. 

\begin{figure}[t]
	\centering
	\begin{minipage}{0.21\textwidth}		
		\includegraphics[width=\textwidth]{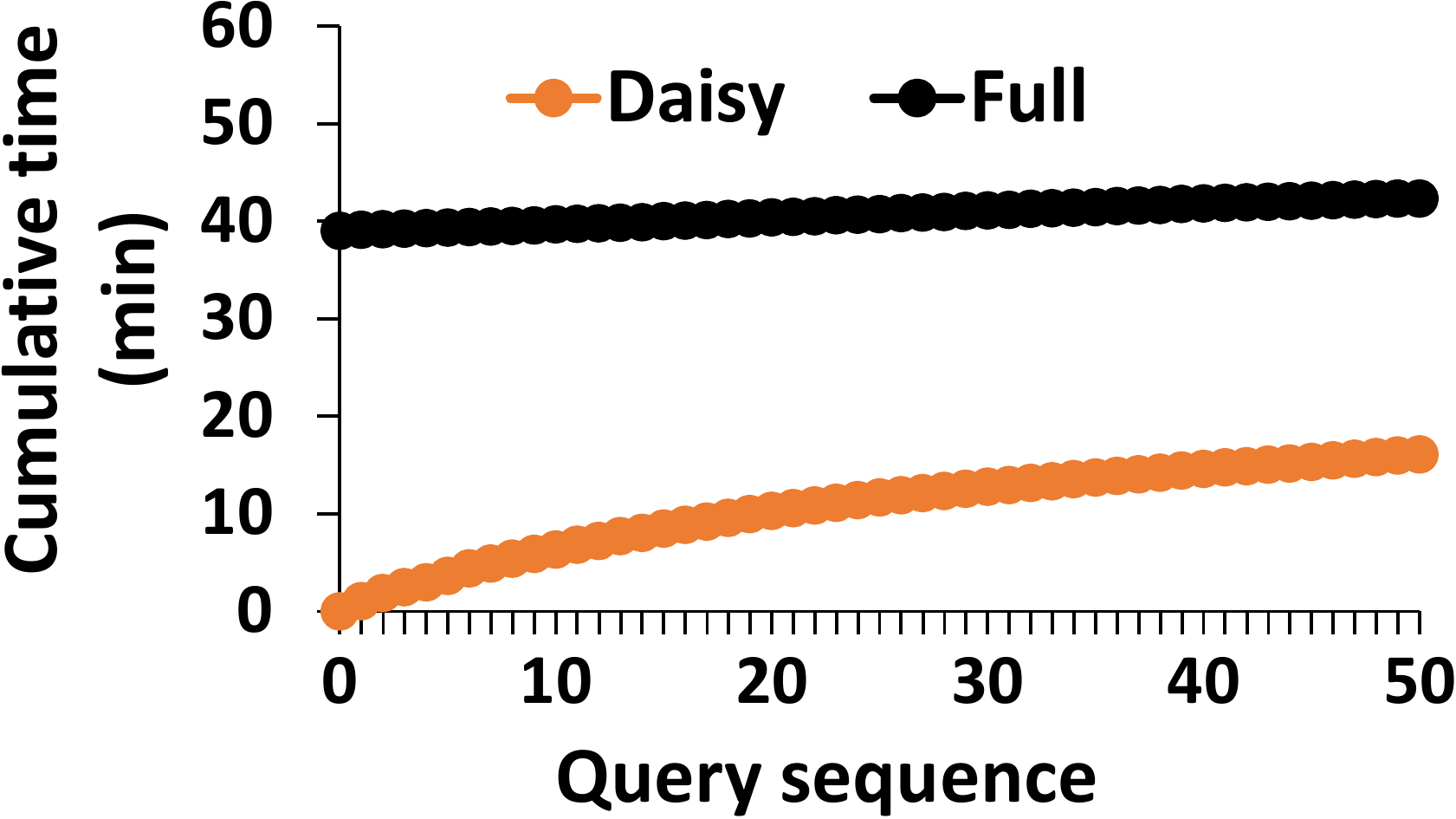}		
		%		\vspace{-1em}	
		\caption{Cost for join queries.}
		\label{fig:join}
	\end{minipage}
	\hspace{2em}
	\begin{minipage}{0.21\textwidth}
		\includegraphics[width=\textwidth]{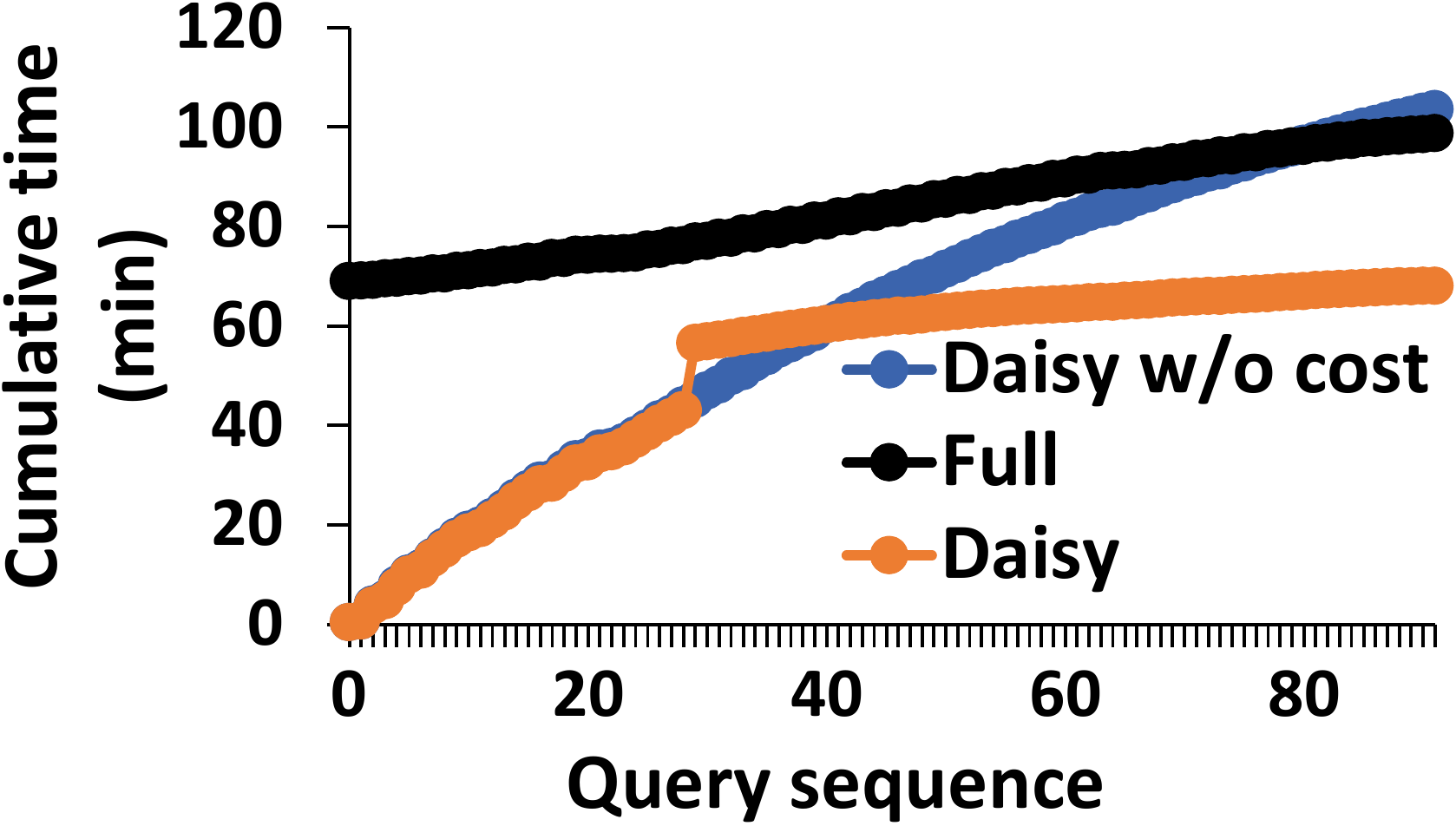}		
		%		\vspace{-1em}	
		\caption{Cost for mixed workload.}
		\label{fig:join_random}
	\end{minipage}
	\vspace{-2em}	
\end{figure}

%%%%%%%%%%%%%%%%%%%ADD%%%%%%%%%%%%%%%%%%%%%%%
Fig. \ref{fig:join} shows the response time of SPJ queries using \system and the offline 
approach respectively. \system outperforms
full cleaning for two reasons: First, similarly to the SP queries case, \system benefits from computing the set of correlated tuples, thereby
restricting the number of comparisons. % for cleaning the erroneous values.
%Therefore, when repairing the erroneous cells, the system employs the extra set of tuples in order to identify
%candidate values, and avoids having to go through the whole dataset.
Second, \system benefits from incrementally updating the join result when extra tuples are added. On the other
hand, offline cleaning performs a probabilistic join which is expensive.
%since there is a filter condition on the $rhs$ of $\phi$ over the lineorder table,
%\system exploits the logical-level optimizations and avoids having to perform a second iteration in order to check 
%for violations in the lineorder table after cleaning the suppliers table. \system can avoid the extra iteration
%because after cleaning the lineorder table,
%the cleaning process has computed all the tuples which are candidates to satisfy the filter condition. Thus, even
%if there is an update in the suppliers table after the cleaning tasks, the extra tuples of the suppliers relation 
%will match with the existing lineorder tuples. As a result there will be no new, undiscovered violations.

Fig. \ref{fig:join_random} shows the time taken to execute a workload with both SP and SPJ queries.
The lineorder table violates rule \textit{$\phi$:orderkey$\rightarrow$suppkey}, and
the suppliers violates rule \textit{$\psi$:address $\rightarrow$suppkey}.
We use the scenario of Fig.~\ref{fig:sp_random}, where we
execute 90 queries over the 100K version of \textit{lineorder}, and the suppkeys contain 500 distinct values.
Both the SP and join queries are non-overlapping, involve equality and
range conditions and have random selectivities.
We observe that \system predicts that it is more efficient to clean the full dataset
after 30 queries, and thus by penalizing some queries, overall it is faster than both
incremental and full cleaning.

%%%%%%%%%%%%%%%%%ADD%%%%

\begin{figure}
	\vspace{1em}
	\includegraphics[width=.24\textwidth]{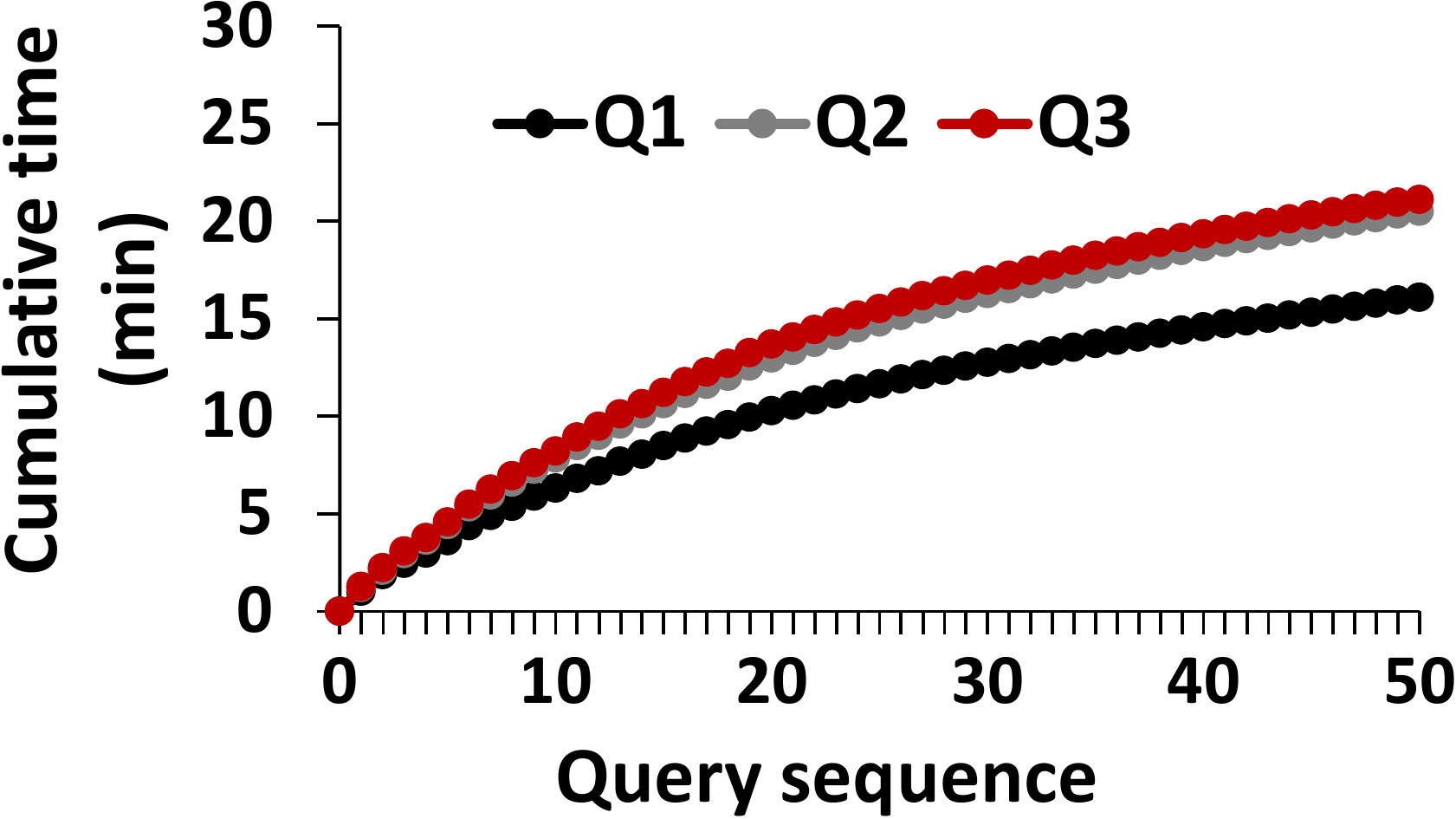}		
	%			\vspace{-1em}	
	\caption{Cost for complex queries of SSB workload.}
	\label{fig:ssb}
	\vspace{-1.5em}
\end{figure}

Fig. \ref{fig:ssb} compares the response time of three query workloads from the SSB family to evaluate how \system behaves with more complex queries.
We use the same setup with Fig. \ref{fig:join}. \textit{Q1} is a join between \textit{lineorder} and \textit{suppliers}
and contains a range filter on the \textit{suppkey}. \textit{Q2} additionally joins the result of \textit{Q1} with \textit{part} and \textit{date} tables and groups by year and brand.
\textit{Q3} contains a fourth join with \textit{customer}. All queries project the keys of the involved tables thus the probabilistic orderkey/suppkey attributes as well.
We observe that regardless of the query complexity, since \system pushes down the cleaning operator, cleaning affects only the join between \textit{lineorder} and \textit{suppliers}.
The breakdown of the cost of the overall plan showed that the time difference between \textit{Q1} and \textit{Q2,Q3} stems from the fact that in \textit{Q2,Q3} the initial join projects
the extra attributes required for the following joins. Thus, cleaning is more expensive since Spark requires outer joins to split and stitch back the clean and dirty part of the query result.

\subsection{Real-world scenarios}
\label{subsec:realworld}

In this set of experiments, we compare \system against Holoclean, and we also measure the cost of executing a realistic exploratory analysis scenario.
In all experiments, \system outputs the same results with the scale-out cleaning approach.

\noindent\textbf{Hospital:}
We evaluate the efficiency and accuracy of \system and Holoclean.
We use rules \textit{$\phi_1$:$\forall$$t_1$,$t_2$:$\urcorner$($t_1$.zip=$t_2$.zip$\wedge$$t_1$.city$\neq$ $t_2$.city)}, 
\textit{$\phi_2$:$\forall$$t_1$,$t_2$:$\urcorner$($t_1$.hospitalName=$t_2$.hospitalName$\wedge$$t_1$.zip $\neq$$t_2$.zip)},
\textit{$\phi_3$:$\forall$$t_1$,$t_2$:$\urcorner$($t_1$.phone=$t_2$.phone $\wedge$$t_1$.zip$\neq$$t_2$.zip)}. 
To obtain a fair comparison, we execute Daisy on a single node, and when measuring response time,
we disable the inference of Holoclean; we obtain only the candidate
values for each cell.
For accuracy, we apply Holoclean's inference using \systemNoindent's domain generation (DaisyH), and we compare it with the original Holoclean.
To integrate \system with Holoclean, we populate the \textit{cell\_domain} table that Holoclean uses with the candidate values that \system computes.
We also report \systemNoindent's accuracy when selecting the most probable value (DaisyP).
%We use three versions of the dataset with 1K, 10K, and 100K tuples respectively. 
For accuracy, we use the 1K version for which master
data exists. For efficiency, we use version 100K.
%We measure the accuracy of an exploratory analysis scenario in which a user learns the rules as a result of exploring through the dataset.
%Table \ref{tab:hospital_acc} presents how the accuracy increases in \system and Holoclean respectively.
%We observe that when not all constraints are known, Holoclean fails to correctly clean the erroneous values because it relies
%on the known constraints. However, \system manages to infer the correct values by keeping multiple values until all the constraints
%are known. 

\begin{figure}[t]
	%%%%%%%%%%%%%%%%%%%%%%%%%REMOVE%%%%%%%%%%%%%%%%%%%%%%%%%%%%%
	%	\begin{minipage}{0.21\textwidth}
	%		\includegraphics[width=\textwidth]{figures/join_random}			
	%		\captionsetup{width=1.6\linewidth}
	%		\vspace{-1.4em}
	%	     \caption{Cost for mixed workload.}
	%		\label{fig:join_random}
	%	\end{minipage}
	%	\hspace{2em}
	\begin{minipage}{0.5\textwidth}	
		\scriptsize
		\begin{tabular}{|l|c|c|c|c|c|c|c|c|c|}
			\hline
			\multirow{3}{*}{\textbf{}} &
			\multicolumn{3}{c}{\textbf{$\phi_1$}} &
			\multicolumn{3}{|c}{\textbf{$\phi_1$+$\phi_2$}} &
			\multicolumn{3}{|c|}{\textbf{$\phi_1$+$\phi_2$+$\phi_3$}} \\			
			\hline			
			& Prec. & Rec. & F1 & Prec. & Rec. & F1& Prec. & Rec. & F1\\
			\hline
			Holoclean & 1  &  0.55 & 0.71& 0.98 & 0.95 & 0.96 & 0.98 & 0.92 & 0.95 \\
			\hline
			DaisyH & 0.97 & 0.52 & 0.68 & 1 & 0.98 & 0.99 & 1 & 0.98 & 0.99 \\ 
			\hline
			DaisyP & 0.41 & 0.51 & 0.45 & 1 & 0.97 & 0.98 & 1 & 0.98 & 0.99 \\ 
			\hline						
		\end{tabular}
		%		  \captionsetup{width=1\linewidth}
		\captionof{table}{Accuracy}
		\label{tab:hospital_acc}			
	\end{minipage}	
	\vspace{-1.7em}		
\end{figure}

Table \ref{tab:hospital_acc} shows the precision, recall, and F1-measure for \system and Holoclean.
\system executes a workload of 4 SP queries that access the whole dataset. Each tuple is accessed only once and is cleaned at query time.
For Holoclean we clean errors offline and measure the accuracy of the corresponding attributes.
We observe that both systems exhibit comparable accuracy. When not all rules are known, such as in the case of $\phi_1$, Holoclean performs better because it generates the domain using quantitative statistics, 
whereas DaisyH uses the correlations driven by the dependencies. DaisyP performs worse because it blindly selects the most probable value.
However, when more rules are known, \system is more accurate because Holoclean
prunes the domain of each value using a threshold
for performance. Hence, using \systemNoindent's optimizations, one can avoid trading accuracy at this level.

Table \ref{tab:hospital_time} shows the response time when we clean violations of different subsets
of $\phi_1$, $\phi_2$, $\phi_3$.
\system outperforms both Holoclean and the full cleaning approach due to the optimizations it enables.
%Also, as we increase the number of rules, the difference between \system and the offline approach is more
%significant since rule $\phi_3$ has more violations thereby incurring a higher overhead to clean and combine the probabilistic values
%stemming from the rules.
Holoclean exhibits higher response times as the tuples of hospital are highly correlated; it performs multiple comparisons to compute the candidate values. Also, Holoclean, traverses multiple times the dataset for each dirty group
to compute the domain. As a result, \systemNoindent's optimizations can be
applied to the domain construction of Holoclean in an analysis-aware scenario. 

%Table \ref{tab:hospital_time_incr} shows the response time when cleaning violations of different rule combinations.
%In the case of \system the user queries the whole dataset and executes the cleaning task.
%For Holoclean, we measure only the cost of the candidate fixes for each cleaning task.
%We measure the cost of checking $\phi_1$, $\phi_1+\phi_2$, and $\phi_1+\phi_2+\phi_3$.
%We compare the scenario where we execute \system and Holoclean three times, one for each rule set, with a single execution of \system that incrementally updates the probabilistic data.
%Holoclean exhibits higher response times since in the case of hospital the tuples are highly correlated, thus for each erroneous cell it requires multiple
%comparisons to compute the candidate values. 
%We observe that the single execution of \system outperforms the three separate executions since it 
%can merge the probabilistic fixes by inducing only the overhead of merging the fixes.
%maintains provenance information and merging violations
%from rules that appear one after the other. 

\begin{table}[t]	
	\centering
	\scriptsize	
	%	\vspace{-0.7em}
	\begin{tabular}{|l|c|c|c|}
		\hline
		\textbf{} & \textbf{$\phi_1$} & \textbf{$\phi_1+\phi_2$} & \textbf{$\phi_1+\phi_2+\phi_3$} \\
		\hline
		Full cleaning & 51 sec & 49 sec & 118 sec \\
		\hline
		\system & 49 sec & 40 sec & 92 sec\\
		\hline
		Holoclean  & 1020 sec & 1108 sec & 1188 sec \\
		\hline
	\end{tabular}
	\caption{Response time when increasing number of rules. }
	\label{tab:hospital_time}
	\vspace{-2em}
\end{table}

\begin{table}[t]	
	\centering
	\scriptsize	
	%	\vspace{-0.7em}
	\begin{tabular}{|l|c|c|c|c|}
		\hline
		\textbf{} & \textbf{$\phi_1$} & \textbf{$\phi_1+\phi_2$} & \textbf{$\phi_1+\phi_2+\phi_3$} & Total\\
		\hline
		\system (3 executions) & 51 sec & 49 sec & 118 sec & 218 sec\\
		\hline
		\system (1 execution)  & 51 sec& 41 sec& 40 sec& 132 sec\\
		\hline
		Holoclean  & 1020 sec & 1108 sec & 1188 sec & 3316 sec\\
		\hline
	\end{tabular}
	\caption{Response time when increasing the number of rules. \system maintains provenance information and updates the probabilistic data based on the new rule without having to execute the task from scratch.}
	\label{tab:hospital_time_incr}
	\vspace{-1.5em}
\end{table}

Table \ref{tab:hospital_time_incr} shows the benefit stemming from maintaining provenance information to the original data
and incrementally updating the probabilistic data in the case new rules appear.
We measure the total cost by checking $\phi_1$, $\phi_1+\phi_2$, and $\phi_1+\phi_2+\phi_3$.
We use a scenario where we execute \system and Holoclean three times, one for each rule set, and compare it with a single execution of \system that incrementally updates the probabilistic data.
We evaluate the case where a user queries the whole dataset and executes the cleaning task, thus the cost of \system is equivalent to the offline cost.
For Holoclean, we measure only the cost of the candidate fixes for each cleaning task.
We observe that the single execution of \system outperforms the three separate executions since it 
can merge the probabilistic fixes by inducing only the overhead of merging the fixes.

%%%%%%%%%%%%%%%%%%%%%%%%%%%%%%%%%%%ADD%%%%%%%%%%%%%%%%%%%%%%%%%%%%%%%%%%%%%%%%%%
\noindent\textbf{Nestle exploratory analysis:}
Data scientists working for Nestle, often need to apply analysis to discover
information about different coffee products. 
We simulate this scenario and execute a query workload of 37 SP queries in which the analyst 
requests the details of a given coffee product through the $Category$ attribute.
The dataset contains violations of the FD $Material \rightarrow Category$. 
$Material$ represents the material out of which each product is made; in the case of coffee
products it represents the type of beans. $Category$ is the type
of product. 

Table \ref{tab:real} shows the response time of the analysis over the two versions of the dataset.
In both cases, the queries access 40\% of the dataset. We observe that in the smaller dataset (20MB), the difference
in the response time stems only from the fact that the analysis accesses 40\% of the dataset. However, when the dataset becomes 
bigger, the difference is more significant. \system is faster because the selectivity of the $Category$ attribute is
very small, and thus since it appears with multiple erroneous $Material$ values, the full cleaning approach ends up
iterating through the dataset multiple times.

%\begin{table}[t]
%	\vspace{-0.7em}
%	\centering
%	\footnotesize
%	\begin{tabular}{|l|c|c|}
%		\hline
%		\textbf{} &  \textbf{$\phi_1$+$\phi_2$} & \textbf{$\phi_1$+$\phi_2$+$\phi_3$}\\
%		\hline
%		\system & 1 & 1\\
%		\hline
%		Holoclean & 0.98  &  0.97\\
%		\hline
%	\end{tabular}
%	\caption{Precision of candidate fixes.}
%	\label{tab:hospital_acc}			
%	\vspace{-1em}
%\end{table}

%\begin{table}[t]	
%	\centering
%	\scriptsize	
%	%	\vspace{-0.7em}
%	\begin{tabular}{|l|c|c|}
%		\hline
%		\textbf{} & \textbf{$\phi_1+\phi_2$} & \textbf{$\phi_1+\phi_2+\phi_3$}\\
%		\hline
%		\system (3 executions) & 1.96x & 3.03x  \\
%		\hline
%		\system (1 execution)  & 1.8x & 2.58x \\
%		\hline
%		Holoclean  & 2x & 3.3x \\
%		\hline
%	\end{tabular}
%	\caption{Slowdown when adding new rules after having executed $\phi_1$. }
%	\label{tab:hospital_time_incr}
%	%	\vspace{-2.5em}
%\end{table}

\noindent\textbf{Air quality exploratory analysis:}
This scenario is similar to the analysis that data scientists perform in Kaggle~\cite{air_quality_analysis}, where they observe 
how air pollution evolves over the years in the US.
Specifically, an analyst checks the CO measurements at specific locations, one location per state, district, or territory.
The query workload consists of 52 queries each of which outputs the average CO measurement for a given county
grouped by year.
%\vspace{-0.3em}
%\begin{minted}[escapeinside=||,fontsize=\footnotesize]{sql}
%SELECT year, AVG(co)
%FROM air_quality
%WHERE county_name="[county]"
%GROUP BY year
%\end{minted}
%\vspace{-0.2em}
Table \ref{tab:real} shows that offline cleaning
is unable to terminate after a timeout of one day due to having to perform multiple iterations for each erroneous group
over a larger dataset in order to clean it.
%since the $county$ and $state$ attributes have small selectivity.

\begin{table}[t]
	\centering
	\scriptsize	
	\begin{tabular}{|l|c|c|}
		\hline
		\textbf{Dataset} & \textbf{\system} & \textbf{Offline}\\
		\hline
		%%%%%%%%%%%%%%%%%%%%%%%%%%%%%%%%%ADD%%%%%%%%%%%%%%%%%%%%
		Nestle (20MB) & 2.9 min & 3.97 min\\
		\hline
		Nestle (200MB)  & 26.8 min  & 8.5 hours\\
		\hline
		Air quality 30\% & 10.5 min &  - \\
		\hline
		Air quality 97\% & 49 min & - \\
		\hline
	\end{tabular}
		\caption{Response time on realistic scenarios. }	
		\label{tab:real}					
	\vspace{-2.4em}
\end{table}

\noindent \textbf{Summary.} The optimizations at the executor level ensure that \system scales better than offline
approaches by restricting the comparisons to clean the data. 
%\system can also clean and execute queries faster 
%by maintaining a partially probabilistic
%dataset, 
%than it computes the full set of probabilities.  
%Moreover, using the
%probabilistic values, \system handles the uncertainty of the input rules, and avoids re-checking the whole
%dataset when the user adds rules while executing queries. 
Moreover, the logical-level optimizations enable \system to configure
the optimal placement of cleaning operators, depending on the query workload and the errors.

\section{Conclusion}
\label{sec:conclusion}

Data scientists usually perform multiple iterations over a dataset in order to understand and prepare it for data analysis.
Having to apply each cleaning task over the whole dataset each time is tedious and time-consuming. 
Having data cleaning decoupled from data analysis also increases human effort as data cleaning is a subjective process that
highly depends on the data analysis that users need to perform.

Our work introduces \systemNoindent, a system that partially cleans the dataset through exploratory queries.
\system integrates cleaning operators inside the query plan, and efficiently executes them over dirty data
by providing probabilistic answers for the erroneous entities.
We evaluate \system using both synthetic and real workloads and show that it scales better than approaches that fully clean the dataset as an offline process.

\noindent\textbf{Acknowledgments.} We would like to thank the reviewers for
their valuable comments and suggestions. This work is partially funded by the EU FP7 programme (ERC-2013-CoG), under grant agreement no 617508 (ViDa) 
and the European Union's Horizon 2020 research and innovation programme under grant agreement no 825041(SmartDataLake).

%\clearpage

\bibliographystyle{ACM-Reference-Format}
\bibliography{references}

\end{document}